\documentstyle[psfig]{l-aa}
\begin{document}
\thesaurus{ }
\def\etal{{\it et al.}}
\def\lae{\mathrel{<\kern-1.0em\lower0.9ex\hbox{$\sim$}}}
\def\gae{\mathrel{>\kern-1.0em\lower0.9ex\hbox{$\sim$}}}

\title {A complete sample of GHz-Peaked-Spectrum 
radio sources and its radio properties
}

\author{
C. Stanghellini\inst{1,5} \and C.P. O'Dea\inst{2} \and D. Dallacasa\inst{4} 
\and S.A. Baum\inst{2} \and R. Fanti\inst{3,4} \and C. Fanti\inst{3,4}
}

\institute{
Istituto di Radioastronomia del CNR, C.P. 141, I-96017 Noto SR, Italy 
\and 
Space Telescope Science Institute, 3700 San Martin Drive, Baltimore, MD, 21218 
\and
Dipartimento di Fisica, Universit\`a degli Studi, via Irnerio 46, I-40126 Bologna, 
Italy.
\and
Istituto di Radioastronomia del CNR, Via Gobetti 101, I-40100 Bologna, Italy 
\and
visitor at the Space Telescope Science Institute, Baltimore
}

\offprints{C. Stanghellini}

\date{ Submitted 18/12/97}
\maketitle

\begin{abstract}

We define a complete sample of thirty-three  GHz-Peaked-Spectrum (GPS) 
radio sources based on their spectral properties.
We present measurements of the radio spectra and polarization of the complete
sample and a list of additional GPS sources which fail  one or more criteria 
to be included in the complete sample.

The majority of the data have been obtained from quasi-simultaneous
multi-frequency observations at the Very Large Array (VLA) during 3 
observing sessions. Low frequency data 
from the Westerbork Synthesis Radio Telescope (WSRT)
and from the literature have been combined with the VLA data 
in order to better define the spectral shape.
 
The objects presented here show a rather wide range of
spectral indices at high and low frequencies, including a few
cases where the spectral index below the turnover is close
to the theoretical value of 2.5 typical of self-absorbed
incoherent synchrotron emission.
Faint and diffuse extended emission is found in about 10\% of the sources.

In the majority of the GPS sources, the fractional polarization is found 
to be very low, consistent with the residual instrumental polarization 
of 0.3 $\%$.
\footnote{Tables 4 and 5 are only
available in electronic form at the CDS via anonymous
ftp to cdsarc.u-strasbg.fr (130.79.128.5)
or via http://cdsweb.u-strasbg.fr/Abstract.html}

\end{abstract}

\section{ Introduction}

The GHz-peaked-spectrum (GPS) radio sources are characterized by a 
simple convex spectrum which peaks in a range of about a decade 
around 1 GHz.  General discussions on the general properties of
these objects are given by O'Dea \etal\  (1991)
O'Dea and Baum (1997), and O'Dea (1998)
where an exhaustive   bibliography can also be found.

Common characteristics of the bright sample of GPS radio sources are: 
small size ($\lae 1$ kpc), high radio luminosity, low fractional 
polarization, and apparently low variability.
They are a mixed class of quasars and
galaxies. Galaxies tend to be $L_*$ or brighter and at redshifts
$0.1 \lae z \lae 1$  (O'Dea \etal\ 1996) while 
quasars are often found at  very large redshift $1 \lae z \lae 4$, 
(O'Dea 1990).

Currently, there are two main competing hypotheses to explain the origin of
GPS radio sources and their possible evolution.

In the ``young source" scenario, first suggested by Phillips and Mutel (1982), 
GPS radio sources with compact double (CD) morphology  or  compact 
symmetric morphology (Compact Symmetric Objects: CSO)
are classical double radio sources at the very first stage of
their lives. 

In the ``frustration'' scenario, GPS radio sources
will never become as large as the classical doubles since
they are confined to the sub-kpc scale by a dense and
turbulent ambient medium (e.g., O'Dea \etal\ 1991, Carvalho 1994, 1998).
However, it is unclear whether the gas density and its distribution in the
nuclear region is sufficient to confine the radio source 
for times of the order of 10$^7$ years (see e.g., O'Dea 1998).

Recent results support the young source  hypothesis. 
Fanti \etal\  (1995)  presented a model for the evolution of  the  
galactic-scale Compact Steep Spectrum (CSS) sources into large scale
classical doubles.  They argued that the objects showing symmetric 
morphology are probably not confined by a dense and clumpy medium 
(see also De Young 1993). They also suggest that
the typical CSS source age is of the order of 10$^6$ years and that
the radio source luminosity decreases by an order of magnitude as the size
of the radio source grows from a few kpc to hundreds of kpc (see also 
Begelman 1996).
The same conclusions are reached by Readhead et al. (1996a,b) in their
study of a few CSO's.  O'Dea and Baum (1997), combining
the complete sample of GPS presented here with the CSS sources by
Fanti \etal\ (1990) reach similar conclusions. They further note
note that the luminosity evolution required makes 
it likely that  GPS and CSS  radio sources will  
evolve into objects less powerful than the most powerful classical doubles.  

The detection of arc-second scale faint extended emission around $\sim 10\%$
of GPS sources  (Baum \etal\ 1990, Stanghellini \etal\ 1990)
motivated Baum \etal\ (1990) to suggest that nuclear activity is recurrent
in these sources. In this hypothesis, we see the relic of a previous epoch 
of activity as faint diffuse emission surrounding the current young nuclear 
source. 

In order to achieve a deeper understanding of the GPS radio sources
we pursued the following project.
 
(1) We created  a complete sample of GPS radio sources by means
of a preliminary bibliographic research, checked with subsequent 
new multi-frequency data from the Very Large Array (VLA) 
and the Westerbork Synthesis Radio Telescope (WSRT).

(2) We determined the properties of the radio spectrum and the polarization.

(3) We obtained optical imaging to determine the host galaxy properties. 

(4) We obtained  VLBI observations to study the  milliarcsecond 
radio morphology.  

In this paper we present the selected sample and the
results from the VLA and WSRT radio observations. The optical properties of
GPS radio sources are discussed by O'Dea, Baum and Morris (1990), 
Stanghellini
\etal\ (1993), and O'Dea \etal\  (1996). The milliarcsecond morphology
is presented in Stanghellini \etal\  (1997). Constraints on radio source
evolution based on our observations of the complete GPS sample are discussed 
by O'Dea \& Baum (1997). 

H$_0$=100 km sec$^{-1}$ Mpc$^{-1}$, and q$_0$=0.5 have been used
in this paper.

\section{ The Complete Sample}

Research on GPS radio sources has increased  
in the past few years and a heterogeneous list of objects
has been accumulated. The list presented by O'Dea \etal\  (1991)
consists of $\sim$ 100 radio sources, while Dallacasa and
 Stanghellini (1990)
collected a larger list of GPS and CSS radio sources and candidates.
To enable statistical analysis of the properties of these
objects we selected a {\bf complete sample} of bright GPS radio sources. 

The selection criteria were:

- declination $\delta > -25^o$

- galactic latitude $|l| > 10^\circ$

- flux density at 5 GHz $S_{5GHz}>1Jy$

- turnover frequency between 0.4 and 6 GHz

- spectral index $\alpha _{thin} > 0.5$ ($S_{\nu}\propto \nu^{-\alpha}$)
in the high frequency, likely optically thin, part of the spectrum. 

We started by selecting GPS candidates from the 1 Jy catalog of 
K\"uhr \etal\ (1981). We cleaned this first ``dirty" sample 
(Stanghellini \etal\ 1990) using our multi-frequency
observations from the VLA and the WSRT presented here, supplemented
with data from the literature. 

The final complete sample consists of the 33 objects
listed in Table 1.
This is the first complete sample of bright GPS objects.
The new radio data permit the proper classification
and the improvement in the estimate of important parameters such as
spectral indices, turnover frequencies,
and polarization properties.

We note the following regarding the sample and possible selection
effects. 

\begin{table*}
\caption{Complete sample: column 1 to 10,  source name,  redshift,
optical identification, magnitude and filter/color, flux density at 5 GHz,
observed and rest frame
peak frequency, references for the optical information
1): O'Dea et al. 1991; 2) de Vries et al. 1995; 3)
Stanghellini et al. 1993; 4) Snellen et al. 1996; 5) Stanghellini et al.
1997; 6) de Vries, private communication; 7) White et al. 1993;
8) Hewitt and Burbidge 1987; 9) Hunter et al. 1993; 10) Heckman et al.
1994; 11) Stickel and Ku\"hr 1993a; 12) O'Dea et al. 1996;
13) Stickel and Ku\"hr 1993b; 14) Junkkarinen et al. 1991. Last
column is a note indicating the radio source with extended emission.}
\begin{flushleft}
\begin{tabular}{ccrrcccccccccc}
\hline
\noalign{\smallskip}
\noalign{\smallskip}
name & z & id. & m & S$_{5GHz}$
  & $\nu_{to}$&$\nu_{rest}$  &
 $\alpha _{low\nu}$ & $\alpha _{high\nu}$ &S$_{max}$&ref.&note&&\\
\noalign{\smallskip}
\hline\noalign{\smallskip}
0019--000&0.305&G  &18.4r&1.10&0.8&1.0&-1.0&1.2&3.47&3,4&&&\\
0108+388&0.669 &G  &22.0r&1.29&3.9&6.5&-2.1&0.9&1.33&1,3&e&&\\
0237--233&2.223 &Q  &16.6V&3.34&1.0&3.2&-1.9&0.7&7.05&1&    &&\\
0248+430&1.316 &Q  &15.5V&1.24&5.2&12.1&-0.4&0.6&1.27&1&e?&&\\
0316+161&...&G  &23.6R&2.91&0.8&1.2&-0.7&1.1&9.55&3&&&\\
0428+205&0.219 &G  &19.3R&2.30&1.0&1.3&-0.6&0.6&4.02&1,3 &&&\\
0457+024&2.384 &Q  &19.4V&1.57&1.9&6.3&-3.5&0.6&1.89&1&&&\\
0500+019&0.583&G&21.0i&1.89&2.0&3.2&-1.6&0.9&2.51&2&&&\\
0710+439&0.518 &G  &18.5i&1.67&1.9&2.8&-1.3&0.6&2.09&1,3&&&\\
0738+313&0.631 &Q  &16.1V&3.66&5.1&8.3&-0.7&0.8&3.82&1 &e&&\\
0742+103&...&S? &23r&3.46&2.8&5.7&-0.7&0.7&4.12&1,5 &&&\\
0743--006&0.994 &Q  &17.5V&2.05&6.0&12.0&-0.6&0.8&2.12&1,5&&&\\
0941--080&0.228&G  &17.9r&1.10&0.5&0.6&-0.3&1.0&3.40&3,6&e?&&\\
1031+567&0.459 &G  &20.2r&1.27&1.3&1.8&-0.7&0.8&1.87&1,3 &&&\\
1117+146&0.362&G &20.1R  &1.00&0.5&0.7&-0.4&0.8&3.89&1,3 &&&\\
1127--145&1.187&Q  &16.9V  &3.82&1.0&2.2&-0.8&0.6&5.80&8&&&\\
1143--245&1.95&Q&18.5V&1.40&2.0&5.9&-2.5&0.7&1.69&1&&&\\
1245--197&1.273 &Q  &20.5V&2.31&0.5&1.2&-0.7&0.9&8.69&1& &&\\
1323+321&0.369 &G  &19.2r&2.35&0.5&0.7&-0.7&0.6&7.03&1,3&&&\\
1345+125&0.122 &G &15.5r&3.03&0.6&0.6&-0.9&0.7&8.86&1,3 &&&\\
1358+624&0.431 &G  &19.8r&1.80&0.5&0.8&-1.5&0.7&6.56&1,3&&&\\
1404+286&0.077 &Sy&14.6r&2.66&4.9&5.3&-1.5&1.6&2.76&1,3&&&\\
1442+101 &3.544 &Q  &17.8V&1.19&0.9&4.1&-1.3&0.9&2.61&1&&&\\
1518+047 &1.296 &Q &22.6r&1.05&0.9&2.1&-0.6&1.3&4.58&1,10&&&\\
1600+335&...&G  &23.2r&2.67&2.6&3.9&-0.2&0.9&3.06&1,3&&&\\
1607+268&0.473 &G  &20.4r&1.71&1.0&1.5&-1.5&1.2&5.44&1,3&&&\\
2008--068&...&G  &21.3R&1.34&1.3&2.0&-0.9&0.8&2.64&1,3&&&\\
2126--158&3.270 &Q&17.3V&1.17&3.9&16.7&-1.3&0.5&1.23&1&&&\\
2128+048&0.99&G  &23.3r&2.02&0.8&1.6&-1.0&0.8&4.93&1,11&&&\\
2134+004&1.936&Q  &16.8V&8.50&5.2&15.4&-1.2&0.7&8.59&8&e&&\\
2210+016&...&G &22.0i&1.05&0.4&0.6&-0.6&1.0&4.51&2&&&\\
2342+821& 0.735 &Q  &20.1r&1.28&0.5&0.8&-0.7&0.9&6.29&1,3&&&\\
2352+495&0.237 &G  &18.4R&1.47&0.7&0.8&-0.2&0.5&2.93&1&&&\\
\noalign{\smallskip}
\hline
\noalign{\smallskip}
\end{tabular}
\end{flushleft}
\end{table*}

\begin{table*}
\caption{Additional objects:
column 1 to 10,  source name,  redshift,
optical identification, magnitude and filter/color, flux density at 5 GHz,
observed and rest frame
peak frequency, references for the optical information, references for the optical information
as table 1. Last
column is a note indicating the radio source with extended emission.}
\begin{flushleft}
\begin{tabular}{ccrrcccccccc}
\hline
\noalign{\smallskip}
\noalign{\smallskip}
name & z & id. & m & S$_{5GHz}$
  & $\nu_{to}$&$\nu_{rest}$  &
 $\alpha _{low\nu}$ & $\alpha _{high\nu}$ &S$_{max}$&ref.&note\\
\noalign{\smallskip}
\hline\noalign{\smallskip}
0026+346&0.6&G  &21.0r&1.31&1.4&2.3&-0.2&0.4&1.90&4&\\
0201+113&3.61&Q  &19.5R&0.837&...&...&-0.6&0.2&...&1,7&\\
0404+768&0.599&G&21.1r&2.91&0.3&0.5&-0.3&0.5&8.15&1,3&\\
0440--003&0.844&Q&19.22V&1.13&0.3&0.6&-0.3&0.1&1.21&8&\\
0528+134&2.07&Q  &19.2r&1.96&1.9&5.74&-0.4&0.1&2.14&3,9&e\\
0552+398&2.365&Q&18.0V&6.31&9.5&32.1&-1.6&0.6&7.80&1&\\
0703+468&...&S  &23.1r&0.640&0.7&1.4&-1.7&1.0&2.37&1,3&\\
0711+356&1.620&Q&17.0V&0.861&1.6&4.1&-0.7&0.7&1.42&1&\\
0904+039&...&E&...&0.216&0.6&1.3&-0.4&1.3&1.04&1,2&\\
0914+114&...&G&20.0r&0.120&0.3&0.7&-0.1&1.6&2.30&3&\\
1543+005&0.550&G&20.0R&1.23&...&...&0.1&0.5&...&1,10&\\
1732+094&...&G&20.7R&0.742&2.3&4.5&-1.1&1.1&1.35&1&\\
2015+657&2.845&Q&19.7R&0.667&...&...&-0.3&0.2&...&1,11&\\
2021+614&0.2266&G&17.9R&2.82&8.4&10.2&-0.2&0.6&3.29&1&\\
2050+364&0.354&G&21.2r&3.28&2.1&2.9&-0.4&0.9&5.82&1,12&\\
2137+209&1.576&Q&19R&0.590&...&...&0.3&0.9&...&1,10&\\
2149+056&0.740&G&20.4R&0.840&2.5&4.3&-1.5&0.7&1.09&1,13&\\
2223+210&1.949&Q&18.2V&1.07&...&...&...&...&...&1,14&e\\
2230+114&1.037&Q&17.3V&4.12&0.5&1.0&-0.7&0.4&8.35&1&\\
2337+264&...&G&20.0i&0.996&2.7&5.5&-0.2&0.6&1.12&2&\\
\noalign{\smallskip}
\hline
\noalign{\smallskip}
\end{tabular}
\end{flushleft}
\end{table*}

Since variability may influence the spectral shape, the fact that GPS
sources are thought to not be significantly variable may be just due 
to selection effects. Even if the spectral shape is constant, if the
flux density varies significantly, the  source would
probably not be recognized as GPS using data from the literature spanning several
years. Therefore it is important to have simultaneous multi-frequency  observations 
to build a sample  unbiased with respect to variability (see also sect. 4.4).

The  turnover frequency and the spectral indices at low and high frequencies  
are difficult to derive when few data are available and/or the spectrum 
bends continuously. For this reason we included objects with a turnover frequency 
in a range a little larger than that of the canonical decade around 1 GHz. 

The selection is based on the {\it observed} turnover frequency, hence we 
select  objects with different intrinsic turnover frequencies at low and high redshifts.  
A selection based on the intrinsic turnover frequency is not currently possible
due to the lack of complete redshift information
on large samples of radio sources.

We note that the sample is unbiased regarding the optical identification.

\section{ The Observations}

The flux densities obtained in the different observing sessions 
are presented in Tables 4 and 5. Selected data from the literature have 
been added  (with the relevant references) when considered compatible 
with our observations and useful to define the spectral shape. 
  
During the VLA observing sessions an additional 20 GPS candidate sources 
were  observed (Table. 2).  These sources fail one or more criteria of the
complete sample. Nevertheless, the majority of them show a clear GPS-like
spectrum.   
 
\subsection{ VLA observations}

The VLA data were obtained in a number of observing sessions 
in the A configuration spanning several years.  
Eighteen candidate GPS radio sources were observed on
30 December 1984. 
The results from this first set of observations were already 
published by O'Dea \etal\ (1990), Stanghellini \etal\  (1990),
and O'Dea \etal\ (1991). We consider here only the observed
sources belonging to the complete sample.

A second set of VLA observations was obtained on 9 February 1990, and
a third session on 8 June 1991. The journal of these
observations in shown in Table 3.  In the third session, where the majority of 
the sources belonging to the complete sample were observed, we split the 2 
frequencies in each band to obtain the maximum possible separation with adequate 
sensitivity.  This was used in the previous observing sessions only in the L band,
here we used it also in the C, X and P bands. 
This technique proved to be very useful, because it permitted a better 
definition of the
spectral shape and a more accurate selection of the sources peaking close
to the limit of 0.4 GHz that we adopted for our sample. 

Furthermore, the observations separated in frequency permit us to 
solve the $n\pi$ ambiguity that occurs in the determination
of the Faraday rotation measure.

\begin{table}
\caption{Journal of the second and third VLA observing sessions.
Only a few sources were observed at 22 GHz}
\begin{tabular}{ccc}
\hline
\noalign{\smallskip}
\noalign{\smallskip}
date & $\nu_{GHz}$ & $\Delta \nu_{MHz}$\\
\noalign{\smallskip}
\hline\noalign{\smallskip}
09 Feb 1990 & 1.380 & 12.5 \\
09 Feb 1990 & 1.630 & 12.5 \\
09 Feb 1990 & 4.815 & 12.5 \\
09 Feb 1990 & 4.865 & 12.5 \\
09 Feb 1990 & 8.435 & 12.5 \\
09 Feb 1990 & 8.485 & 12.5 \\
09 Feb 1990 & 22.435 & 50 \\
09 Feb 1990 & 22.485 & 50 \\
08 Jun 1991 & 0.302 & 3 \\
08 Jun 1991 & 0.333 & 3 \\
08 Jun 1991 & 1.335 & 12.5 \\
08 Jun 1991 & 1.665 & 12.5 \\
08 Jun 1991 & 4.535 & 12.5 \\
08 Jun 1991 & 4.985 & 12.5 \\
08 Jun 1991 & 8.085 & 12.5 \\
08 Jun 1991 & 8.465 & 12.5 \\
\noalign{\smallskip}
\hline
\noalign{\smallskip}
\end{tabular}
\end{table}

\begin{table}
\caption{Flux densities for the complete sample, see Table 5 for notes and references}
\begin{flushleft}
\begin{tabular}{ccrrc}
\hline
\noalign{\smallskip}
\noalign{\smallskip}
source&$\nu_{GHz}$  &S$_{Jy}$&err.&ref.\\
\noalign{\smallskip}
\hline\noalign{\smallskip}
        $0019-000$ & 0.325 & 2.37 & .24  & 4 \\     
         &0.365  &2.87  &0.04  &31\\
         &0.408  &2.99  &0.14  &8\\
         &0.606  &3.44  &0.20  & 9 \\              
         &0.750  &3.34  &0.16  & 9 \\       
         &1.380  &2.88  &0.06  & 1 \\          
         &1.640  &2.59  &0.06  & 1 \\        
         &4.860  &1.10  &0.02  & 1 \\         
        &22.460  &0.175  &.018 & 1 \\        
\noalign{\smallskip}
\hline
\noalign{\smallskip}
        $0108+38$  
        &0.325	&0.065 &.003& 11\\ 
        &0.608 &0.075 &.002& 10 \\
        &1.380 &0.427 &.01  & 2\\
        &1.630 &0.572 &.02  & 2\\
&2.3 &0.837 &0.03 & 6\\
        &2.695 &1.00 & 0.02  & 9\\
&3.9& 1.26& 0.04&  6\\
        &4.816 &1.30 &0.04   &2\\
        &4.866 &1.29 &0.04   &2\\
&7.7 &0.96 &0.03  &6\\
        &8.434 &0.938 &0.03  & 2\\
        &8.484 &0.933 &0.03   &2\\
&11.2& 0.66& 0.02& 6\\
        &22.460 &0.384 &0.03  & 2\\
\noalign{\smallskip}
\hline
\noalign{\smallskip}
        $0237-233$ 
        &0.365&   2.96& 0.06 & 7  \\ 
        &0.408&   3.67 & 0.11  & 8  \\      
        &0.468&   4.76 & 0.12  & 9  \\   
        &0.635&   6.20 & 0.25  & 9  \\     
        &0.960&   7.02 & 0.20  & 9  \\       
        &1.380&  6.19  &0.12  & 1  \\        
        &1.640&  5.91  &0.12  & 1 \\         
        &4.860&  3.34  &0.07  & 1 \\             
        &8.870&  2.36  &0.08  & 9 \\
        &22.460&  1.09 & 0.10 &  1\\
\noalign{\smallskip}
\hline
\noalign{\smallskip}
        $0248+430$
        &0.302& 0.452 &0.015  &3\\
        &0.333& 0.481 &0.015  &3\\
        &1.335& 0.804 &0.03   &3\\
        &1.665& 0.922 &0.03   &3 \\
        &4.535& 1.244 &0.04   &3\\
        &4.985& 1.241 &0.04   &3\\
        &8.085& 1.13  &0.03   &3\\
        &8.465& 1.09  &0.03   &3\\
        &43.0 & 0.385 &0.01  &14 \\
\noalign{\smallskip}
\hline
\noalign{\smallskip}
       $0316+161$ 
        &0.318 & 6.80 &  0.30  &9\\        
        &0.365 & 7.55&   0.11 &7\\
        &0.408 & 8.06 &  0.25  &8  \\            
        &0.468 & 8.91 &  0.22  &9    \\        
        &0.635 & 9.64 &  0.24  &9 \\                 
        &0.750 & 9.54 &  0.24  &9 \\          
        &0.96 & 9.40 &  0.30   &6 \\       
        &1.380	&7.83  & 0.20   &2\\
        &1.630	&7.24  & 0.20   &2  \\
        &2.3    &5.34  & 0.20  &6\\
        &3.9    &3.61  & 0.03  &6\\
        &4.815	&2.96	&0.10  &2\\
        &4.865	&2.91& 0.10 & 2\\
        &7.7    &1.79& 0.08 & 6\\
        &8.435	&1.70& 0.07 &2 \\
        &8.485	&1.69& 0.07 &2  \\
        &11.2   &1.16& 0.03 &6  \\
        &22.460 &0.570&  0.030&  2\\
\noalign{\smallskip}
\hline
\noalign{\smallskip}
\end{tabular}
\end{flushleft}
\end{table}
\addtocounter{table}{-1}

\begin{table}
\caption{continued}
\begin{flushleft}
\begin{tabular}{ccrrc}
\hline
\noalign{\smallskip}
\noalign{\smallskip}
source&$\nu_{GHz}$  &S$_{Jy}$&err.&ref.\\
\noalign{\smallskip}
\hline\noalign{\smallskip}
        $0428+205$ 
        &0.325 &2.63  &0.08  &4\\
        &0.408 &2.92  &0.15  &5\\
        &0.635 &3.96  &0.27  &9\\
        &0.968 &4.07  &0.13  &9\\
        &1.380 &3.70  &0.10  &2\\
        &1.630 &3.62  &0.10  &2\\
        &4.815 &2.32  &0.07  &2\\
        &4.865 &2.30  &0.07  &2\\
        &8.435 &1.71  &0.05  &2\\   
        &8.485 &1.71  &0.05  &2\\   
        &22.460 &0.96 & 0.05  &2\\
\noalign{\smallskip}
\hline
\noalign{\smallskip}
$0457+024$ 
        & 0.333	 &0.16   &0.02 &  3\\
        & 0.408  &0.33   &0.01 &  5\\         
        & 0.96  &1.40   &0.06&  6\\
        & 1.335  &1.74  &0.06 & 3  \\     
        & 1.665  &1.84  &0.06 & 3  \\      
        & 2.3  &1.97   &0.06 & 6 \\
        & 4.535  &1.65  &0.05 & 3  \\       
        &  4.985 & 1.57 & 0.05&  3 \\             
        & 8.085  &1.19  &0.04  &3\\
        & 8.465  &1.15  &0.04  &3\\
\noalign{\smallskip}
\hline
\noalign{\smallskip}
 $0500+019$ 
&0.325 &0.471& .025&  4\\
          &0.365& 0.628& 0.029& 7\\
          &0.606& 1.31&  0.09 & 9\\
          &0.96 & 1.81&   0.10 & 6\\
          &1.380& 2.23&  0.07 & 1\\
          &1.640& 2.31&  0.07 & 1\\
          &2.3  & 2.26&  0.08 & 6\\
          &3.9  & 2.15&  0.03 & 6\\
          &4.860& 1.89&  0.06 & 1\\
          &8.870& 1.35&  0.05 & 9\\
          &11.2 & 1.23&  0.03 & 6\\
          &43.0   & 0.33&  .02 & 14\\
\noalign{\smallskip}
\hline
\noalign{\smallskip}
$0710+439$ 
        &  0.325&  0.56&  0.03&  4\\      
        & 0.365&  0.655&  0.019& 7 \\  
        & 0.408&  0.75 &  0.04 & 15\\        
        & 0.608&  1.27 &  0.06 & 4 \\         
        & 1.380&  1.94 & 0.06  &2  \\        
        & 1.630&  1.95 & 0.06  &2  \\          
        & 4.816&  1.68 & 0.05  &2  \\        
        & 4.866&  1.67 & 0.05  &2  \\        
        & 8.434&  1.28 & 0.04  &2  \\             
        & 8.484&  1.28 & 0.04  &2  \\               
        & 22.460&  0.66&   0.03&  2 \\ 
\noalign{\smallskip}
\hline
\noalign{\smallskip}
 $0738+313$ 
          &0.302& 0.64  &0.04&  3\\
          &0.333& 0.73  &0.04&  3\\
          &0.608& 1.15   &0.04&  4\\
          &1.335& 1.93   &0.06&  3\\  
          &1.665& 2.30   &0.08&  3\\
          &4.535& 3.67   &0.12&  3\\
          &4.985& 3.66   &0.12&  3\\
          &8.085& 3.45   &0.11&  3\\
          &8.465& 3.41   &0.11&  3\\
          &43.&   0.943  &0.03&  14\\  
\noalign{\smallskip}
\hline
\noalign{\smallskip}
\end{tabular}
\end{flushleft}
\end{table}
\addtocounter{table}{-1}

\begin{table}
\caption{continued}
\begin{flushleft}
\begin{tabular}{ccrrc}
\hline
\noalign{\smallskip}
\noalign{\smallskip}
source&$\nu_{GHz}$  &S$_{Jy}$&err.&ref.\\
\noalign{\smallskip}
\hline\noalign{\smallskip}
$0742+103$ 
         &0.302&  1.23   &0.06 & 3\\     
         &0.333&  1.32   &0.06 & 3 \\            
         &1.335&  3.31  &0.10  & 3 \\             
         &1.665&  3.69  &0.10  & 3 \\             
         &2.695&  3.92   &0.20 & 9 \\        
         &4.535&  3.61  &0.10   &3  \\         
         &4.985&  3.46  &0.10   &3  \\         
         &8.085&  2.76  &0.08  &3  \\          
         &8.465&  2.69  &0.08  &3  \\            
         &20.0&  1.44  &0.15  &19\\
         &37.0&  1.00   &0.09  &16\\
 \noalign{\smallskip}
\hline
\noalign{\smallskip}
$0743-006$ 
         &0.302&  0.31 & 0.02&  3\\
         &0.333&  0.38 & 0.02&  3\\
         & 1.335& 0.70 & 0.02&  3\\
         & 1.665& 0.81 & 0.02&  3\\
         & 2.700&  1.40 & 0.08&  9\\
         & 4.535& 1.99  &0.06&  3\\
         & 4.985& 2.05  &0.04&  3\\
         & 8.085& 1.97  &0.06&  3\\
         & 8.465& 1.93  &0.06&  3\\
         & 20.0 &  0.99 & 0.10&   19\\
\noalign{\smallskip}
\hline
\noalign{\smallskip}
$0941-080$ 
         &0.302&  3.23&   0.10 &  3\\      
         &0.333&  3.33&   0.10 &  3\\      
         &0.96 &  3.16&   0.10&  6\\
         &1.335&  2.71&   0.09&  3\\         
         &1.665&  2.43&   0.08&  3\\       
         &2.3  &  1.90&   0.06&  6\\
         &3.9  &  1.41&   0.05&  6\\
         &4.535&  1.20&   0.04&  3\\       
         &4.985&  1.10&   0.04&  3\\       
         &7.7  &  0.738&  0.02&  6\\
         &8.085&  0.691& 0.023&  3\\
         &8.465&  0.657& 0.022&  3\\
        &11.2  &  0.550& 0.02& 6\\
\noalign{\smallskip}
\hline
\noalign{\smallskip}
 $1031+567$ 
          &0.151& 0.78&  0.04 & 27\\  
          &0.325& 1.36&  0.07 & 4\\
          &0.408& 1.50&  0.07 & 5\\
          &0.608& 1.59&  0.08 & 4\\
          &1.380& 1.78&  0.05 & 2\\
          &1.630& 1.78&  0.05 & 2\\
          &4.816& 1.28&  0.04 & 2\\
          &4.866& 1.27&  0.04 & 2\\
          &8.434& 0.82&  0.02&  2\\
          &8.484& 0.81&  0.02&  2\\
\noalign{\smallskip}
\hline
\noalign{\smallskip}
 $1117+146$ 
         &0.302&  3.58 & 0.20  & 3\\      
         &0.333&  3.73 & 0.20  & 3\\              
         &0.365&  3.82 & 0.08& 7\\              
         &0.96 &  2.96  & 0.11 & 6\\
         &1.335&  2.46  & 0.08 & 3\\               
         &1.665&  2.21  & 0.07 & 3\\               
         &2.3  &  1.67  & 0.06 & 6\\
         &3.9  &  1.24  & 0.04 & 6\\
         &4.535&  1.02  &0.03  &3\\               
         &4.985&  1.00  &0.03  &3\\               
         &7.7  &  0.70 &0.02  &6\\   
         &8.085&  0.642 &0.023 & 3\\               
         &8.465&  0.613 &0.022 & 3 \\              
         &11.2 &  0.51 &0.02  &6 \\   
\noalign{\smallskip}
\hline
\noalign{\smallskip}
\end{tabular}
\end{flushleft}
\end{table}
\addtocounter{table}{-1}

\begin{table}
\caption{continued}
\begin{flushleft}
\begin{tabular}{ccrrc}
\hline
\noalign{\smallskip}
\noalign{\smallskip}
source&$\nu_{GHz}$  &S$_{Jy}$&err.&ref.\\
\noalign{\smallskip}
\hline\noalign{\smallskip}
 $1127-145$ 
         &0.302 &4.17&  0.20&   3\\
         &0.333 &4.51&  0.20&   3\\
         &1.335 &5.57 & 0.18&  3\\
         &1.665 &5.44 & 0.17&  3\\
         &4.535 &4.01 & 0.13&  3\\
         &4.985 &3.82 & 0.13&  3\\
         &8.085 &3.10 & 0.10&  3\\
         &8.465 &3.03 & 0.10&  3\\
         &14.940& 2.09& 0.30&  7\\
         &20.0 &1.77 & 0.18&  19\\
\noalign{\smallskip}
\hline
\noalign{\smallskip}
$1143-245$ 
         &0.302&  0.175&  0.01&  3\\      
         &0.333&  0.224&  0.01&  3  \\          
         &1.335&  1.518&  0.08&  3\\         
         &1.665&  1.654&  0.08&  3\\          
         &4.535&  1.47 & 0.07 & 3\\          
         &4.985&  1.40 & 0.07 & 3\\          
         &8.085&  1.03 & 0.05 & 3\\            
         &8.465&  0.99&  0.05&  3\\       
\noalign{\smallskip}
\hline
\noalign{\smallskip}
$1245-197$ 
          &0.302& 7.80 & 0.40 &  3\\
          &0.325& 7.86 & 0.30 &  4\\  
          &0.333& 8.22 & 0.40 &  3\\
          &0.365& 8.89& 0.11& 7\\
          &0.408& 8.61&  0.16&  8\\
          &1.335& 5.21&  0.15&  3\\
          &1.665& 4.63&  0.15&  3\\
          &4.535& 2.48&  0.08&  3\\
          &4.985& 2.31&  0.07&  3\\
          &8.085& 1.51&  0.05&  3\\
          &8.465& 1.45&  0.05&  3\\
\noalign{\smallskip}
\hline
\noalign{\smallskip}
$1323+321$ 
         &0.151&  3.78&   0.10 & 26\\    
         &0.318&  6.32&   0.18 & 18 \\     
         &0.408&  7.02&   0.35 & 5 \\         
         &0.608&  6.47&   0.30  & 4 \\      
         &1.380&  4.88&  0.15  &2  \\       
         &1.630&  4.45&  0.15  &2  \\        
         &4.816&  2.35&   0.08 & 2 \\         
         &4.866&  2.35&   0.08 & 2\\  
         &8.435&  1.63&   0.05 & 2\\
         &8.485&  1.63&   0.05 & 2\\        
\noalign{\smallskip}
\hline
\noalign{\smallskip}
 $1345+125$ 
          &0.178& 4.60  & 0.46 & 17\\   
          &0.302& 7.86 & 0.30  & 3\\
          &0.333& 8.39 & 0.30  & 3\\
          &0.365& 8.31& 0.17& 7\\
          &0.408& 8.78 & 0.27&  8\\
          &1.335& 5.33 & 0.15&  3\\
          &1.665& 4.84 & 0.15&  3\\
          &4.535& 3.18 & 0.10&  3\\
          &4.985& 3.03 & 0.10&  3\\
          &8.085& 2.28 & 0.07&  3\\
          &8.465& 2.21 & 0.07&  3\\
          &20.0& 1.13 & 0.11&  19\\
\noalign{\smallskip}
\hline
\noalign{\smallskip}
\end{tabular}
\end{flushleft}
\end{table}
\addtocounter{table}{-1}

\begin{table}
\caption{continued}
\begin{flushleft}
\begin{tabular}{ccrrc}
\hline
\noalign{\smallskip}
\noalign{\smallskip}
source&$\nu_{GHz}$  &S$_{Jy}$&err.&ref.\\
\noalign{\smallskip}
\hline\noalign{\smallskip}
$1358+624$ 
         &0.151&  1.71 & 0.04 & 27\\     
         &0.302&  4.79 & 0.15 & 3   \\     
         &0.333&  5.24 & 0.17 & 3 \\       
         &0.365&  5.56 & 0.07& 7\\
         &0.608&  5.90  & 0.20 & 4\\       
         &1.335&  4.49  & 0.15 & 3\\           
         &1.665&  3.88  & 0.13 & 3\\          
         &4.535&  1.94  & 0.07 & 3\\            
         &4.985&  1.80  & 0.06 & 3\\             
         &8.085&  1.25  & 0.04 & 3\\           
         &8.465&  1.20  & 0.04 & 3\\     
\noalign{\smallskip}
\hline
\noalign{\smallskip}
 $1404+286$ 
          &0.327& 0.177 &0.005 & 12\\
          &0.365& 0.179 &0.032 &7\\
          &0.610& 0.245 &.020  &11\\
          &1.380& 0.79 & 0.03 & 2\\
          &1.630& 1.02  &0.03  &2\\
          &2.300& 1.65  &0.05  &6\\
          &2.695& 2.00  &0.06  &23\\
          &3.900& 2.52  &0.07  &6\\
          &4.816& 2.64  &0.08  &2\\
          &4.866& 2.66  &0.08  &2\\
          &8.434& 1.95  &0.06  &2\\
          &8.484& 1.94  &0.06  &2\\
         &11.200& 1.48  &0.05  &6\\
         &22.460& 0.490 &0.025 &2\\
\noalign{\smallskip}
\hline
\noalign{\smallskip}
$1442+101$ 
         &0.302&  1.61&  0.05 & 3\\      
         &0.333&  1.82&  0.06 & 3 \\      
         &0.608&  2.24&   0.10 &  4\\         
         &0.96 &  2.53&  0.09 &  6\\
         &1.335&  2.50&  0.08 & 3 \\              
         &1.665&  2.35&  0.08 & 3 \\              
         &2.3  &  1.83&  0.07 & 6 \\
         &3.9  &  1.34&  0.04 & 6 \\
         &4.535&  1.30&  0.04 & 3 \\        
         &4.985&  1.19&  0.04 & 3 \\              
         &7.7  &  0.70& 0.02 & 6 \\
         &8.085&  0.76&  0.03&  3\\            
         &8.465&  0.72&  0.03&  3 \\
         &11.2 &0.462 &0.02 &6\\      
\noalign{\smallskip}
\hline
\noalign{\smallskip}
$1518+047$ 
         &0.178& 2.20  & 0.44  &17\\
         & 0.365& 3.404& 0.139& 7\\
         & 0.608& 4.05 & 0.20 & 4\\   
         & 0.96 & 4.72 & 0.24 & 31\\
         & 1.380& 3.98 & 0.12 & 2\\
         & 1.640& 3.53 & 0.11 & 2\\
         & 3.9  & 1.46 & 0.05 & 31\\
         & 4.815& 1.06 & 0.03 & 2\\
         & 4.865& 1.05 & 0.03 & 2\\
         & 8.435& 0.478& 0.015& 2\\  
         & 8.485& 0.480& 0.015& 2\\    
         & 11.1 & 0.305& 0.03 &31\\
\noalign{\smallskip}
\hline
\noalign{\smallskip}
$1600+335$ 
         &0.151&  2.17  & 0.10 & 26\\  
         &0.302&  2.46 & 0.12 & 3 \\       
         &0.333&  2.45 & 0.12 & 3 \\        
         &0.365&  2.47 & 0.03& 7\\
         &1.335&  3.05  & 0.10&  3\\         
         &1.665&  3.11  & 0.10&  3\\         
         &4.535&  2.75  & 0.09&  3\\           
         &4.985&  2.67  & 0.09&  3\\         
         &8.085&  2.22  & 0.07&  3\\         
         &8.465&  2.17  & 0.07&  3\\              
         &20.0&  1.02  &0.10 & 19\\    
\noalign{\smallskip}
\hline
\noalign{\smallskip}
\end{tabular}
\end{flushleft}
\end{table}
\addtocounter{table}{-1}

\begin{table}
\caption{continued}
\begin{flushleft}
\begin{tabular}{ccrrc}
\hline
\noalign{\smallskip}
\noalign{\smallskip}
source&$\nu_{GHz}$  &S$_{Jy}$&err.&ref.\\
\noalign{\smallskip}
\hline\noalign{\smallskip}
 $1607+268$ 
          &0.318 & 2.02  &0.08 & 18\\
          &0.365 & 2.53 &0.07& 7\\
          &0.408 & 2.95  &0.15 &  5\\
          &0.635 & 4.51  &0.28 &  9\\
          &0.750 & 5.16  &0.18 &  9\\
          &1.380 & 4.86 &0.15  & 2\\
          &1.630 & 4.42 &0.14  & 2\\
          & 4.816&  1.72& 0.05 &  2\\
          &4.866 & 1.71 &0.05  & 2\\
          &8.434 & 0.96& 0.03 &  2\\
          &8.484 & 0.96& 0.03 &  2\\
         &22.460 & 0.293& 0.015&  2\\
\noalign{\smallskip}
\hline
\noalign{\smallskip}
 $2008-068$ 
          &0.325 & 1.05 &  0.20&  4\\      
          &0.365 & 1.35 & 0.04 &7 \\  
          &0.608 & 1.95 &  0.20&  4\\       
          &1.380 & 2.62 & 0.08 & 2 \\        
          &1.630 & 2.48 & 0.08 & 2 \\      
          &4.816 & 1.34 & 0.04 & 2 \\         
          &4.866 & 1.34 & 0.04 & 2 \\        
          &8.434 & 0.84&  0.05&  2\\            
          &8.484 & 0.78&  0.07&  2\\
          &22.460& 0.37&  0.02&  2\\
\noalign{\smallskip}
\hline
\noalign{\smallskip}
  $2126-158$ 
          &0.333  &0.097 &0.010&  3\\
          &1.335  &0.559 &0.018&  3\\
          &1.665  &0.680 &0.022&  3\\
          &4.535  &1.18  &0.04 &  3\\
          &4.985  &1.17  &0.04 &  3\\
          &8.085  &1.03  &0.03 &  3\\
          &8.465  &1.01  &0.03 &  3\\
          &90.000 &0.31  &0.03 & 13\\
\noalign{\smallskip}
\hline
\noalign{\smallskip}
$2128+048$ 
         &0.302  &3.01 & 0.15 & 3\\      
         &0.333  &3.56 & 0.15 & 3\\       
         &0.365  &3.67 & 0.05& 7\\
         &0.408  &4.11  & 0.12 & 8\\          
         &0.635  &5.10  & 0.23 & 9\\          
         & 0.95  &4.95  & 0.25 & 6\\          
         &1.335  &4.08  & 0.13 & 3\\            
         &1.665  &3.73  & 0.12 & 3\\           
         &2.650  &3.16  & 0.02 & 9\\           
         & 3.9   &2.50  & 0.12 & 6\\
         &4.535  &2.16  & 0.07 & 3\\             
         &4.985  &2.02  & 0.07 & 3\\
         &8.085  &1.37  & 0.04 & 3\\
         &8.465  &1.31  & 0.04 & 3\\
         &14.940 &0.82  & 0.12 & 7\\
         &31.400 &0.43  & 0.11 & 9\\    
\noalign{\smallskip}
\hline
\noalign{\smallskip}
\end{tabular}
\end{flushleft}
\end{table}
\addtocounter{table}{-1}

\begin{table}
\caption{continued}
\begin{flushleft}
\begin{tabular}{ccrrc}
\hline
\noalign{\smallskip}
\noalign{\smallskip}
source&$\nu_{GHz}$  &S$_{Jy}$&err.&ref.\\
\noalign{\smallskip}
\hline\noalign{\smallskip}
 $2134+004$ 
          &0.302&  0.60& 0.03 &  3\\
          &0.333&  0.66& 0.03 &  3\\
          &1.335&  3.25 & 0.11 &  3\\ 
          &1.665&  4.62 & 0.15 &  3\\
          &4.535&  8.61 &  0.30 &  3\\
          &4.985&  8.50 &  0.30 &  3\\
          &8.085&  7.40 & 0.25 &  3\\
          &8.465&  7.25 & 0.20 &  3\\
          &20.0&  4.99  &0.50  &19\\
          &43.0&  2.92  &0.12  &14\\
          &87.0&   1.74 & 0.32 & 20\\
\noalign{\smallskip}
\hline
\noalign{\smallskip}
$2210+016$ 
         &0.302 & 4.23 & 0.20  & 3\\      
         &0.333 & 4.53 & 0.20  & 3\\         
         &0.365 & 4.54 & 0.06 &7\\
         &0.408 & 4.65 &  0.21 & 8\\        
         &0.96  & 3.17 & 0.16  & 6\\
         &1.335 & 2.84 &  0.10 & 3\\         
         &1.665 & 2.45 &  0.08 & 3\\          
         &2.3   & 1.77 &  0.05 & 6\\
         &3.9   & 1.26 &  0.04 & 6\\
         &4.535 & 1.15 &  0.04 & 3\\            
         &4.985 & 1.05 &  0.03 & 3\\          
         &7.7   & 0.66&  0.03 & 6\\
         &8.085 & 0.65&  0.02 & 3\\             
         &8.465 & 0.62&  0.02 & 3\\
         &11.2  & 0.45&  0.02 & 6\\
\noalign{\smallskip}
\hline
\noalign{\smallskip}
 $2342+821$ 
          &0.151& 3.64&  0.10 &  25\\ 
          &0.302& 5.77&  0.30 &  3\\
          &0.325& 5.90&  0.30 &  4\\
          &0.333& 5.91&  0.30 &  3\\
          &0.609& 5.70&  0.30 &  4\\
          &1.335& 3.90&  0.13&  3\\
          &1.665& 3.28&  0.11&  3\\
          &4.535& 1.41 & 0.05&  3\\
          &4.985& 1.28 & 0.04&  3\\
\noalign{\smallskip}
\hline
\noalign{\smallskip}
$2352+495$ 
         &0.151&  1.45&   0.15 &30\\
         &0.325&  2.50&   0.12 & 4\\
         &0.408&  2.61&   0.22 & 14\\
         &0.608&  2.82&   0.06 & 10\\
         &0.966&  2.66&   0.27 & 9\\
         &1.380&  2.59&  0.08 & 2\\
         &1.630&  2.43&  0.07 & 2\\
         &4.816&  1.48&  0.05 & 2\\
         &4.866&  1.47&  0.05 & 2\\
         &8.434&  1.07&  0.03 & 2\\
         &8.484&  1.06&  0.03 & 2\\
        &22.460&  0.683&  0.035& 2\\
\noalign{\smallskip}
\hline
\noalign{\smallskip}
\end{tabular}
\end{flushleft}
\end{table}

\begin{table}
\caption{Flux densities for the additional objects.References to the flux density measures
for tables 4 and 5: 1)VLA first session
2)VLA second session; 3)VLA third session; 4)WSRT;
5)Northern Cross; 6)Ratan 600; 7)Douglas et al. 1996; 8)Large et al.
1981; 9) K\"uhr et al. (1981);
10) Baum et al 1990; 11) de Bruyn private comunication; 12) de Bruyn, 1990;
13) Steppe et al. 1988;  14) Chandler 1995;
15) Ficarra et al. 1985;
16) Esko private communication ;
17) Gower et al. 1967, Pilkington et al. 1965; 
18) Dennison et al. 1981; 19) Edelson 1987;
20) Ter\"asranta et al 1992;
23) Fiedler et al. 1987; 24) Steppe et al. 1992;
25) Baldwin et al. 1985;  
26) Hales et al. 1988;   
27) Hales et al. 1990; 
28) Hales et al. 1991; 
29) Hales et al. 1993a;  
30) Hales et al. 1993b;
31) Khabrakhmanov private communication.
{\bf notes:} the errors are taken as 3\% if they were found
in literature to be less then 3\%.
0108+388: the flux density at 608 MHz and higher frequencies
refers to the core only, while the flux density measure at 325 MHz
likely includes the extended structure found close to the main compact
component (Baum et al. 1990).
0552+398, the flux density at 37 GHz is the average of
the data taken in 1990 (Ter\"asranta et al. 1992), 
and the error is the r.m.s of the distribution. 
The flux density at 90GHz is the average of
the data taken in 1990 (Steppe et al. 1992), and the error
is the r.m.s of the distribution.
2134+004  The flux density at 87 GHz is the average of
the data taken between 1985 and 1988 (Ter\"asranta et al. 1992), 
the error is the r.m.s of the distribution.}
\begin{flushleft}
\begin{tabular}{ccrrc}
\hline
\noalign{\smallskip}
\noalign{\smallskip}
source&$\nu_{GHz}$  &S$_{Jy}$&err.&ref.\\
\noalign{\smallskip}
\hline\noalign{\smallskip}
\noalign{\smallskip}
$0026+346$  
         &0.365 & 1.36&    0.05&    7\\
         &0.408	&1.88	&0.18	&9\\
         &1.380	&1.91	&0.06	&2\\
         &1.630	&1.87	&0.06	&2\\
         &4.815 & 1.32&	0.04&	2\\
         &4.865	&1.31	&0.04	&2\\
         &8.435	&1.05	&0.03	&2\\
         &8.485	&1.04	&0.03	&2\\
        &22.460  &0.745 &  0.04  &   2\\
\noalign{\smallskip}
\noalign{\smallskip}
\hline
\noalign{\smallskip}
\noalign{\smallskip}
$0201+113$  
         &0.323  &0.350  & 0.01 &  21\\
         &1.380  &0.852  & 0.03 &  2\\
         &1.630  &0.891  & 0.03 &  2\\
         &4.815  &0.844  & 0.03 &  2\\
         &4.865  &0.837  & 0.03 &  2\\
         &8.435  &0.803  & 0.02 &  2\\
         &8.485  &0.801  & 0.02 &  2\\
         &43.0   & 0.565 &  0.02&   32\\         
\noalign{\smallskip}
\hline
\noalign{\smallskip}
\end{tabular}
\end{flushleft}
\end{table}
\addtocounter{table}{-1}

\begin{table}
\caption{continued}
\begin{flushleft}
\begin{tabular}{ccrrc}
\hline
\noalign{\smallskip}
\noalign{\smallskip}
source&$\nu_{GHz}$  &S$_{Jy}$&err.&ref.\\
\noalign{\smallskip}
\hline
\noalign{\smallskip}
$0404+768$  
         &0.151	&7.12	&0.30	&28 \\
         &0.302	&8.50	&0.40	&3\\
         &0.333	&8.12	&0.40	&3\\
         &0.608 & 7.22&    0.40 &     4\\  
         &1.335 & 5.71&    0.17&     3\\
         &1.665&  5.13&    0.16&     3\\
         &4.535 & 3.08&    0.10&     3\\
         &4.985&  2.91&    0.10&     3\\
         &15.0	&1.60	&0.08&	23\\
\noalign{\smallskip}
\noalign{\smallskip}
\hline
\noalign{\smallskip}
\noalign{\smallskip}
$0440-003$
& 0.302&  1.20&   .06&     3\\
& 0.333&  1.23&   .06&     3\\
& 1.325&  1.15&   .03&     3\\
& 1.665&  1.19&   .03&     3\\
& 4.535&  1.13&   .03&     3\\
& 4.985&  1.13&   .03&     3\\
& 8.085&  1.08&   .03&     3\\
& 8.465&  1.07&   .03&     3\\
\noalign{\smallskip}
\noalign{\smallskip}
\hline
\noalign{\smallskip}
\noalign{\smallskip}
$0528+134$  
& 0.325& 1.13&  0.05&        4 \\ 
& 1.335& 2.08&  0.06&       3\\
& 1.665& 2.09&  0.06&       3\\
& 4.535& 1.97&	0.06&     3\\
& 4.985& 1.96&	0.06&     3\\
& 8.085& 1.84&	0.05&     3\\
& 8.465& 1.82&	0.05&      3\\
\noalign{\smallskip}
\hline\noalign{\smallskip}
$0552+398$   
          &1.380& 1.47	&0.05&  2\\
          &1.630&  1.91& 0.06&  2\\
          &4.815&  6.27& 0.20&   2\\
          &4.865&  6.31& 0.20&  2\\
          &8.435&  7.47& 0.20&  2\\
          &8.485& 7.44	&0.20 &  2\\
          &22.460& 7.01&   0.35&     2\\
          &37.0&  5.06 &   0.20 &     20\\
          &43.0&  4.57 &   0.18&     14\\
          &90.0&  2.88 &   0.45&    24\\  
\noalign{\smallskip}
\hline
\noalign{\smallskip}
$0703+468$ 
          &0.151& 0.41&   0.02      30\\  
          &0.365& 1.77&   0.02  &  7\\
          &0.408& 1.90&    0.04&     15\\
          &1.380&	1.57	&0.05	&2\\
          &1.630&	1.40	&0.05	&2\\
          &4.815&	0.646&	0.02&	2\\
          &4.865&	0.640	&0.02	&2\\
          &8.435&	0.370	&0.01	&2\\
          &8.485& 0.377	&0.01	&2\\
         &22.460& 0.138  &  0.01&     2\\
\noalign{\smallskip}
\hline
\noalign{\smallskip}
$0711+356$    
         &0.151   &     0.42   &   0.04 &  30\\   
         &  0.325 &       0.61 &   0.02  &  4\\
         &   0.608&        1.10&    0.04 &   4\\
         &  1.380&	1.35	&0.05 &   2\\
         &  1.630&	1.43	&0.05	&2\\
         &  4.815&	0.87&	0.03	&2\\
         &  4.865&	0.86&	0.03	&2\\
         &  8.435&	0.64&	0.02	&2\\
         &  8.485& 	0.64&	0.02	&2\\
         & 22.460&        0.290 &  0.015&    2\\ 
\noalign{\smallskip}
\hline
\noalign{\smallskip}
\end{tabular}
\end{flushleft}
\end{table}
\addtocounter{table}{-1}

\begin{table}
\caption{continued}
\begin{flushleft}
\begin{tabular}{ccrrc}
\hline
\noalign{\smallskip}
\noalign{\smallskip}
source&$\nu_{GHz}$  &S$_{Jy}$&err.&ref.\\
\noalign{\smallskip}
\hline
\noalign{\smallskip}
$0904+039$   
&0.365  0.958&   0.033&   & 7\\
 &      0.408&  1.00&    0.06 &    8\\
          &1.380&  0.77&   0.03 &    2\\
          &1.630&  0.68&   0.02 &    2\\
          &4.815&  0.220&   0.007&    2\\
          &4.865&  0.216&   0.007&    2 \\
          &8.435&  0.109&   0.003&    2\\
          &8.485&  0.108&   0.003&    2\\
\noalign{\smallskip}
\hline
\noalign{\smallskip}
$0914+114$  
          &0.365&  2.28 &   0.06 &   7\\
          &0.408&  2.31 &   0.11 &    8\\
          &1.380&  0.76&   0.02 &    2\\
          &1.630&  0.61&   0.02 &    2\\
          &4.815&  0.121&   0.004&    2\\
          &4.865&  0.120&   0.004&    2 \\
          &8.435&  0.050&   0.002&    2\\
          &8.485&  0.049&   0.002&    2\\
\noalign{\smallskip}
\hline\noalign{\smallskip}
$1543+005$  
          &0.365&  2.28&    0.09&    7\\
          &1.380&  2.03&    0.06&     2\\
          &1.630&  1.90&    0.06&     2\\
          &4.815&  1.24&    0.04&     2\\
          &4.865&  1.23&    0.04&     2\\
          &8.435&  0.95&   0.03&     2\\
          &8.485&  0.95&   0.03&     2\\
         &22.460&  0.59&   0.03&     2\\
\noalign{\smallskip}
\hline
\noalign{\smallskip}
$1732+094$   
           &0.365&  0.246  & 0.033  &  7\\
           &1.380&  1.11   & 0.04   &  2\\
           &1.630&  1.22   & 0.04   &  2\\
           &4.815&  0.74  & 0.03   &  2\\
           &4.865&  0.74  & 0.03   &  2\\
           &8.435&  0.490  & 0.015  &  2\\
           &8.485&  0.490  & 0.015  &  2\\
          &22.460&  0.165  & 0.010   &  2\\
\noalign{\smallskip}
\hline
\noalign{\smallskip}
$2015+657$  
          & 0.151&        0.70&    0.07&    29\\ 
          & 0.365&        0.88&  0.06&     7 \\      
          & 1.380&	0.65&   0.02&     2\\
          & 1.630&	0.67&   0.02&	2\\
          & 4.815&	0.67&   0.02&	2\\
          & 4.865&	0.67&   0.02&	2\\
          & 8.435&	0.59&	0.02&	2 \\
          & 8.485&	0.59&   0.02&	2 \\
\noalign{\smallskip}
\hline
\noalign{\smallskip}
$2021+614$   
          &1.380 & 2.11&  0.06  &  2\\
          &1.640 & 2.17&  0.06  &  2\\
          &4.815 & 2.82&   0.08 &  2\\
          &4.865 & 2.82&   0.08 &  2\\
          &8.435 & 3.22&   0.10  &  2\\
          &8.485 & 3.21&   0.10  &  2\\
         &43.0   & 1.24&   0.19 &  14\\
\noalign{\smallskip}
\hline
\noalign{\smallskip}
\end{tabular}
\end{flushleft}
\end{table}
\addtocounter{table}{-1}

\begin{table}
\caption{continued}
\begin{flushleft}
\begin{tabular}{ccrrc}
\hline
\noalign{\smallskip}
\noalign{\smallskip}
source&$\nu_{GHz}$  &S$_{Jy}$&err.&ref.\\
\noalign{\smallskip}
\hline
\noalign{\smallskip}
$2050+364$ 
          &0.365& 3.05&    0.03 &   31\\
          &1.380&	5.18&    0.20 &     2\\
          &1.630&	5.43&    0.20	&2\\
          &4.815&	3.30&    0.10	&2\\
          &4.865&	3.28&    0.10	&2\\
          &8.435&	2.06&    0.06&	2 \\
          &8.485&	2.04&    0.06&	2\\
\noalign{\smallskip}
\hline
\noalign{\smallskip}
 $2137+209$   
           &0.365 &       1.90&    0.03&    7\\
           & 1.380&	1.32   & 0.04&	2\\
           &1.630&	1.21    &0.04 &    2\\
           &4.815&	0.59    &0.02	&2\\
           &4.865&	0.59   &0.02 &    2\\
           &8.435&	0.38   &0.01	&2 \\
           &8.485&	0.38   &0.01     &2\\ 
          &22.460    &    0.160&   0.010&     2\\
\noalign{\smallskip}
\hline\noalign{\smallskip}
\noalign{\smallskip}
\noalign{\smallskip}
$2149+056$   
          & 0.333&    0.091&   0.01 &    3\\
          & 1.335&    0.730&   0.02 &    3\\   
          & 1.665&    0.836&   0.03 &    3\\
          & 4.535&	0.87&	0.030&	3\\
          & 4.985&	0.84&	0.027&	3\\
          & 8.085&	0.614&	0.020&	3\\
          & 8.465& 	0.593&	0.020&	3\\
\noalign{\smallskip}
\noalign{\smallskip}
\noalign{\smallskip}
\hline
\noalign{\smallskip}
\noalign{\smallskip}
\noalign{\smallskip}
$2223+210$  
       &0.302	   & 3.52&   0.20	&  3\\
       &0.333	   & 3.63&   0.20	&  3\\
       &  1.335  &   1.67&     0.05&   3\\     
       &  1.665  &   1.41 &    0.05&   3  \\  
       &   4.535 &    1.03&   0.03&    3 \\ 
       &    4.985&    1.07  &  0.03&    3 \\ 
\noalign{\smallskip}
\noalign{\smallskip}
\noalign{\smallskip}
\hline
\noalign{\smallskip}
\noalign{\smallskip}
\noalign{\smallskip}
$2230+114$   
          &0.302&	7.34&	0.20	&3\\
          &0.333&	7.86&	0.20	&3\\
          &0.365& 8.22&   0.09   & 7\\
          &1.335& 6.44&    0.20  &    3\\  
          &1.665& 6.10&    0.18 &    3  \\
          &4.535&	4.29&	0.12	&3\\
          &4.985&	4.12&	0.12	&3\\
          &8.085&	3.40&	0.10	&3\\
          &8.465&	3.36&	0.10	&3\\
\noalign{\smallskip}
\noalign{\smallskip}
\noalign{\smallskip}
\hline
\noalign{\smallskip}
\noalign{\smallskip}
\noalign{\smallskip}
$2337+264$   
          &1.380&	1.05 &   0.03	&2\\
          &1.630&	1.08 &   0.03 &    2\\
          &4.815&	0.995&   0.03	&2\\
          &4.865&	0.996&   0.03 &    2\\
          &8.435&	0.776&   0.03	&2 \\
          &8.485&	0.760&   0.03 &    2\\ 
         &22.460& 0.423&   .02&     2\\
\noalign{\smallskip}
\noalign{\smallskip}
\hline
\noalign{\smallskip}
\noalign{\smallskip}
\end{tabular}
\end{flushleft}
\end{table}

The data reduction has been performed in a uniform way for all the VLA data
sets. The errors in the flux densities are dominated by the
calibration errors which are estimated to be around 3$\%$ at L, C, X bands,
around 5$\%$ at P and K bands, respectively.

At VLA resolution almost all the objects observed are dominated by a single 
point-like component.  The side lobes have been deconvolved from the images 
using the AIPS implementation of the Clark CLEAN algorithm 
(Clark 1980, Cornwell and Braun 1988). The data have been self-calibrated in
phases (Schwab 1980, Cornwell and Fomalont 1988) with an initial point 
 model, and when subsequent iterations of self-cal have been considered necessary,
we used an appropriate number of non negative components obtained from the images
 in an interactive process until convergence to an acceptable solution was achieved. 
In most cases an amplitude self-calibration has been performed on the data to 
improve the final images and to allow a search for possible extended emission.

\subsection{ The WSRT data}

Westerbork Synthesis Radio Telescope (WSRT) filler observations 
on most of the objects of the complete sample have been obtained during 
1990 at 327 MHz and during 1991 at 608 MHz and 327 MHz.
Each source has been observed in several snapshots a few tens of minutes
long.

The data reduction for the data obtained in the 1990 has been done with the
package DWARF. The data of the 1991 observations have been 
reduced with AIPS. These latter data in a format readable by AIPS
did not have the redundant baselines, therefore the data were not 
self-calibrated, producing a lower dynamic range 
compared with the capabilities of this instrument.

\subsection{Other data}

A few objects were observed during the spring and the summer of 1991
with the Northern Cross, a radio telescope near Bologna operating at 408 MHz 
on the principle of the Mill's cross (Braccesi \etal\  (1969) and Ficarra \etal\  
(1985)). 
The radio telescope dates back to the sixties, however, recent
mechanical and electronics upgrades allow completely automatic observations 
with increased sensitivity.
The flux density values have been set to the scale of Baars (Baars \etal\ 1977)
adopting a flux density of 37.7 Jy for the calibrator source 3C380 (Riley 1988).

\begin{figure*}
     \vbox{\psfig{figure=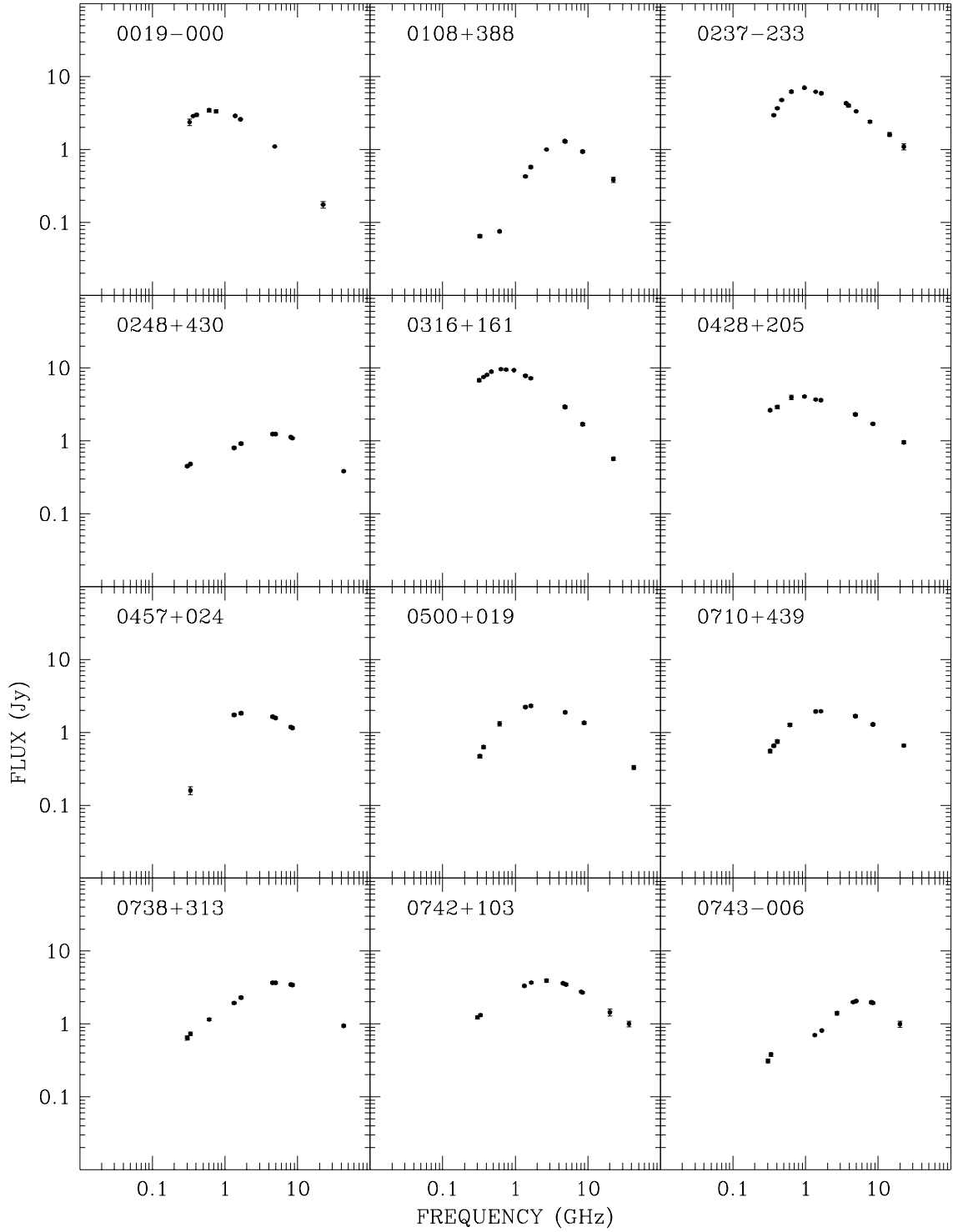,width=16.5cm,height=22cm}}\par
\caption[]{Radio spectra of the radio sources of the complete sample}
\end{figure*}

\begin{figure*}
     \vbox{\psfig{figure=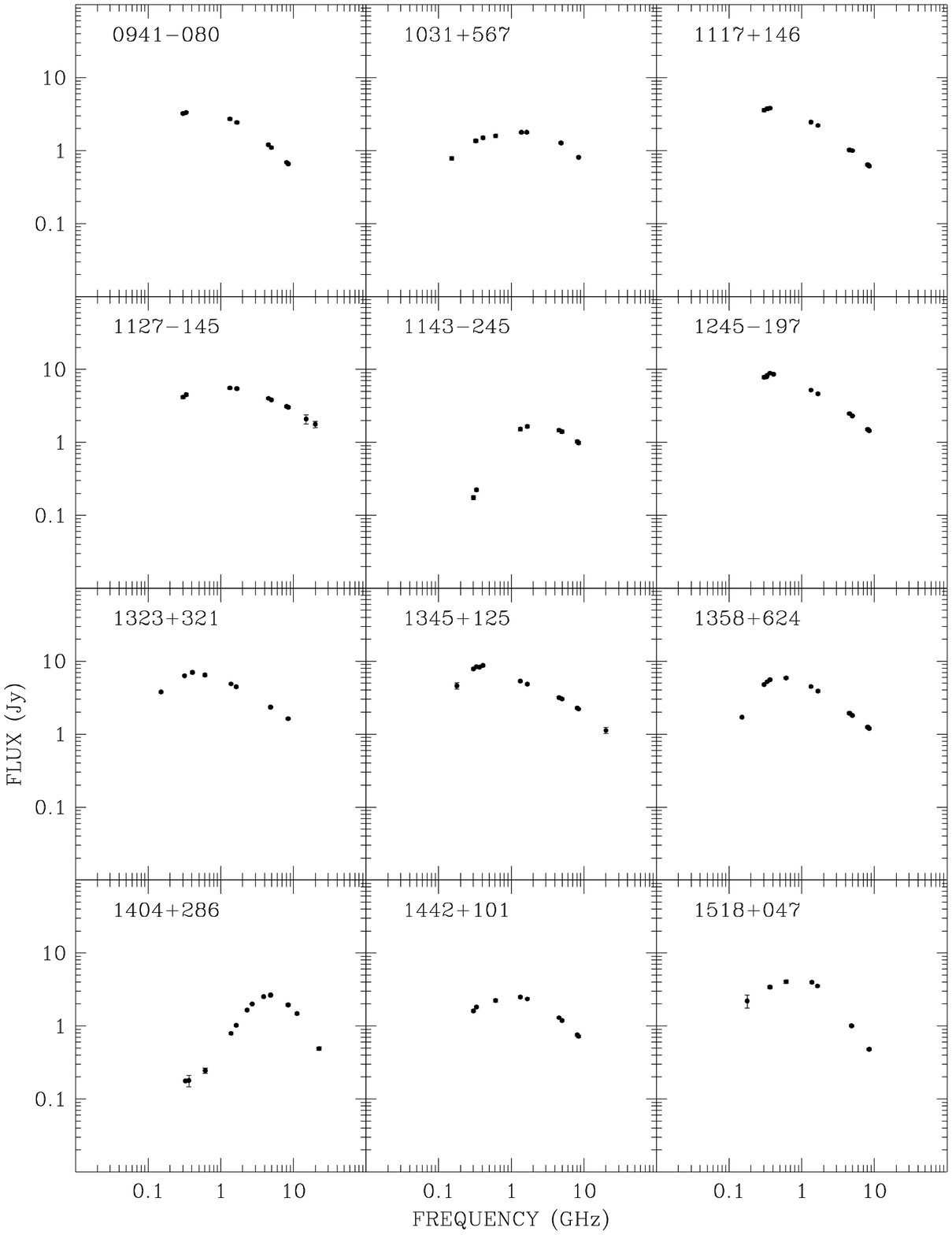,width=16.5cm,height=22cm}}\par
\caption[]{Radio spectra of the radio sources of the complete sample}
\end{figure*}

\begin{figure*}
     \vbox{\psfig{figure=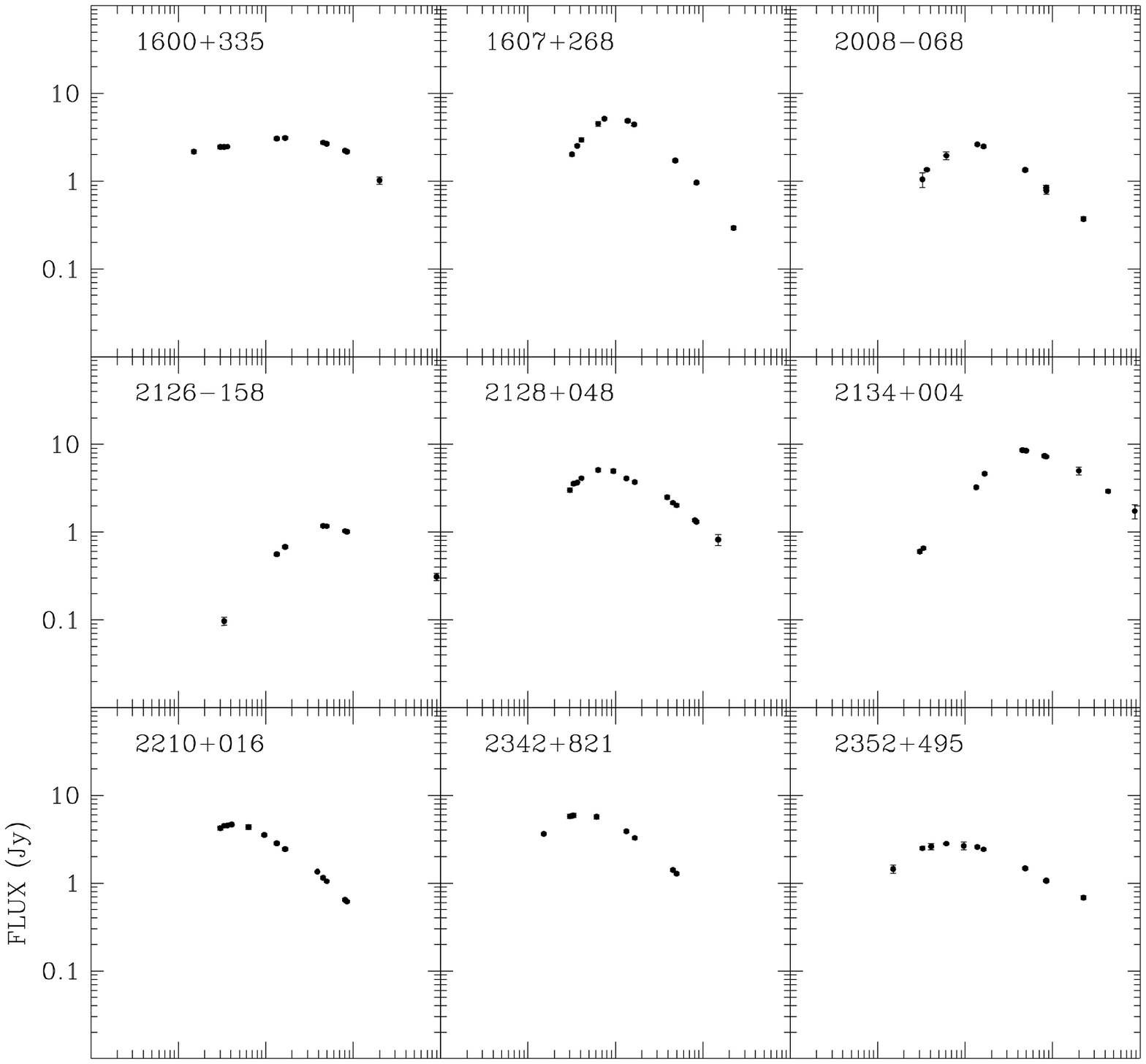,width=16.5cm,height=22cm}}\par
\vskip -6cm
\caption[]{Radio spectra of the radio sources of the complete sample}
\end{figure*}

\begin{figure*}
     \vbox{\psfig{figure=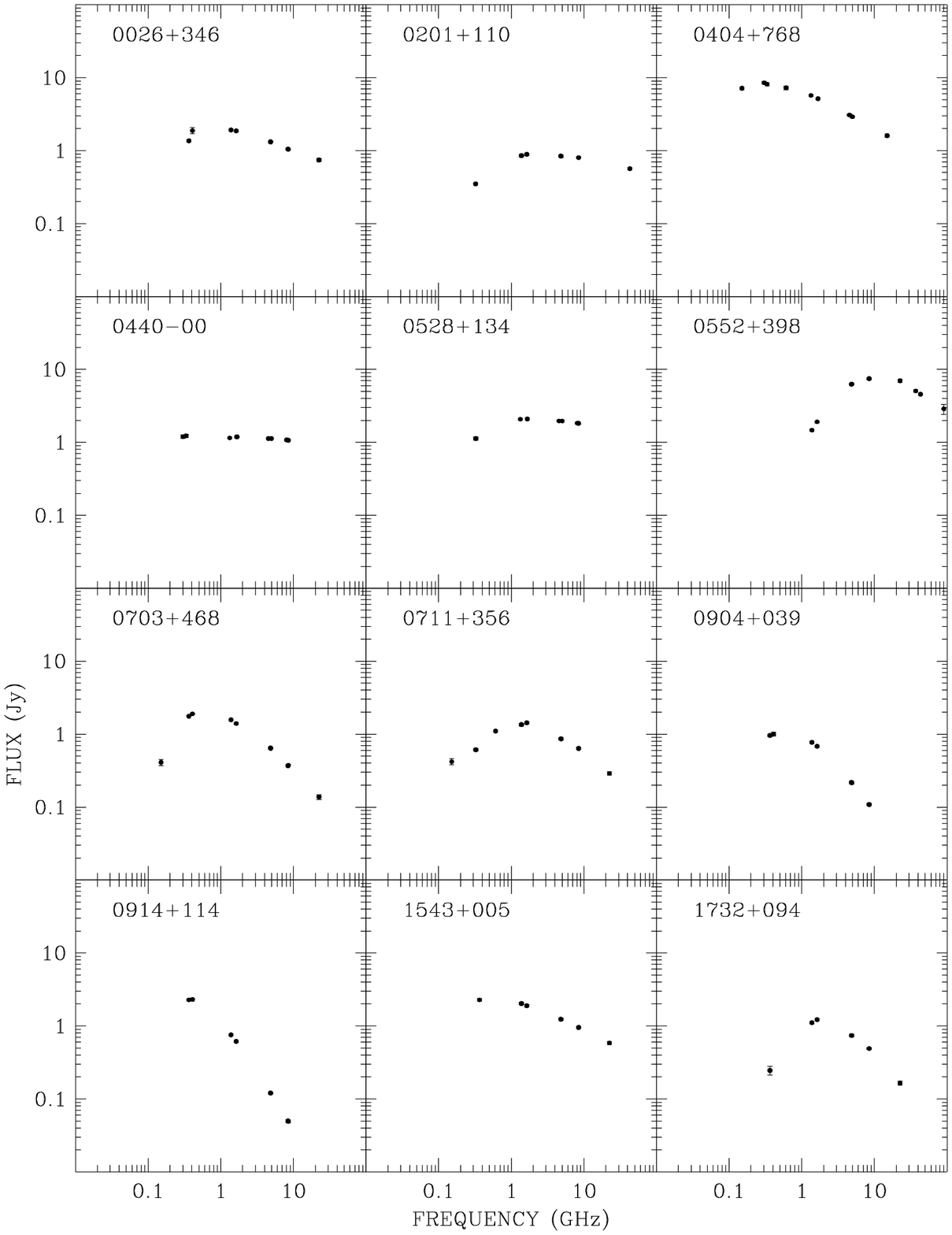,width=16.5cm,height=22cm}}\par
\caption[]{radio spectra of the additional radio sources}
\end{figure*}

\begin{figure*}
     \vbox{\psfig{figure=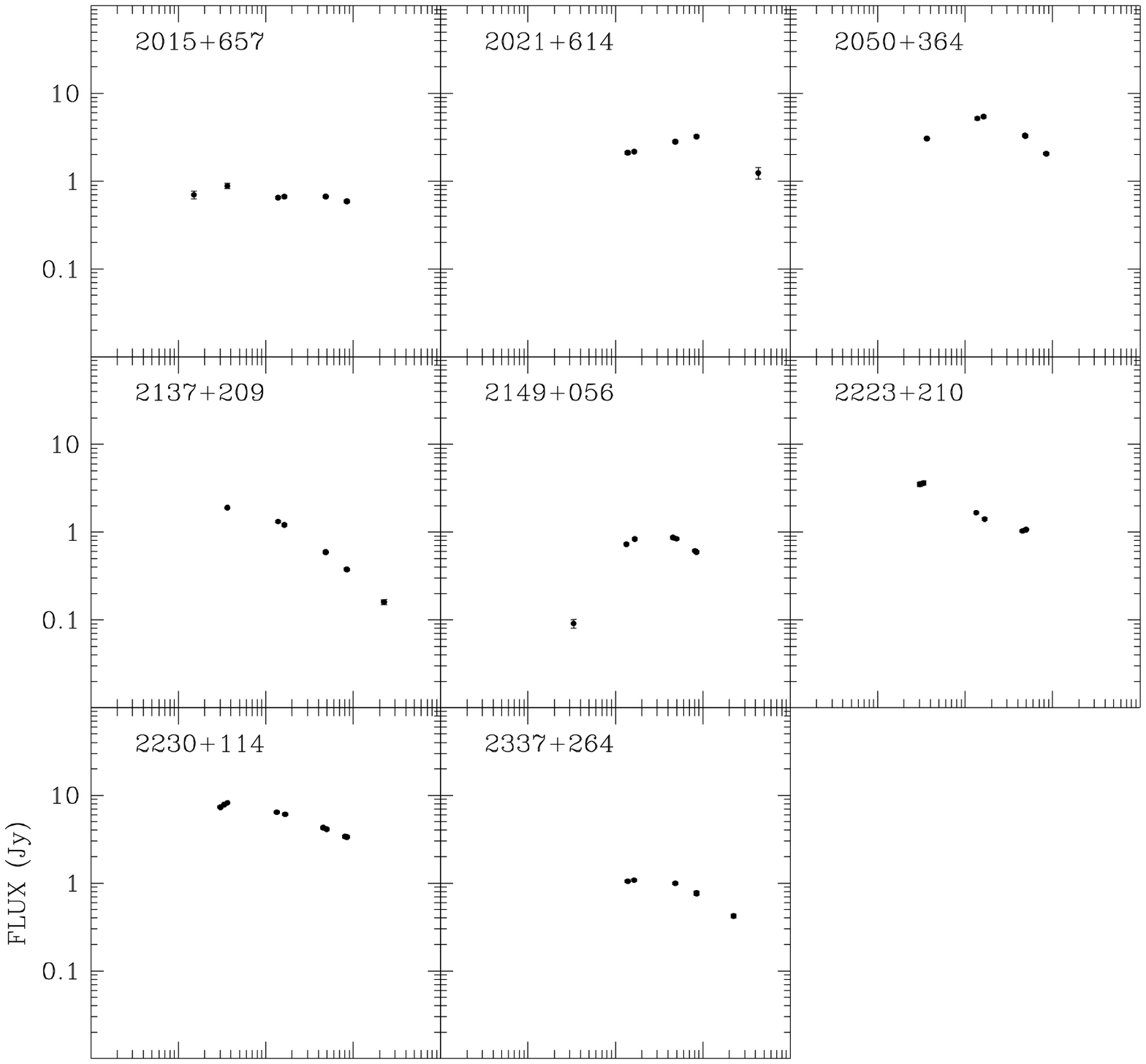,width=16.5cm,height=22cm}}\par
\vskip -6cm
\caption[]{radio spectra of the additional radio sources}
\end{figure*}

\begin{figure*}
     \vbox{\psfig{figure=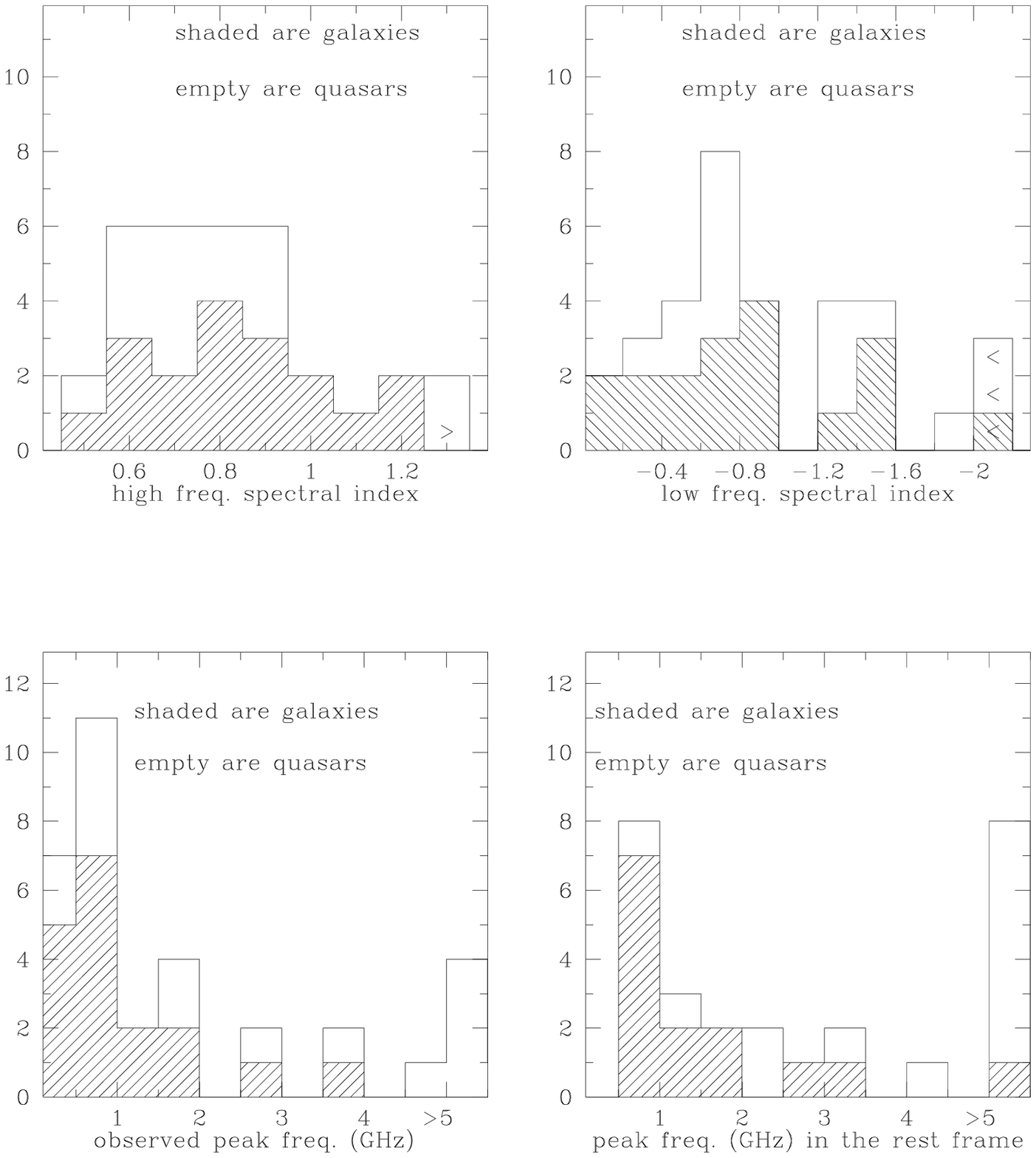,width=16.5cm,height=22cm}}\par
\vskip -6cm
\caption[]{Histograms for the complete sample:
a) high frequency spectral index distribution; b)
low frequency spectral index distribution; c) observed turnover frequency
distribution; d) rest frame turnover frequency distribution }
\end{figure*}

A few additional objects have been observed with the Russian radio telescope 
RATAN 600 (Esepkina \etal\ 1979), in transit mode, during various sessions 
in 1993 and 1994, at 11.2, 7.7, 3.9, 2.3 and 0.96 GHz (Mingaliev, private 
communication). 
 
\section{Results and discussion}

In this paper we present the observational results and the primary 
qualitative conclusions. 
In a future paper we will present a quantitative  discussion of the results.

\subsection{The radio spectra}

The flux densities are presented in Tables 4 and 5 and plotted
in Figures 1 to 5.

In order to extract information from the radio spectra, we 
fitted them with a hyperbola (which is a curve tending asymptotically 
to a straight line at the extrema). We first fit the spectral indices 
in the thick and thin part of the spectrum independently, 
and then fixed  these two values (which correspond to the angular coefficients 
of the asymptotes) and we performed a least square fit,
 solving for the other parameters of the hyperbola.

We calculated the frequency of the spectral peak (Tables 1 and 2) 
using  the fitted curves.  We show the distributions of the observed and rest frame
turnover frequency for the 33 sources of the complete sample in Figures 6c and 6d.
In Fig. 6a and 6b we show  the high frequency (above the peak) and low frequency 
(below the peak) spectral index distribution, respectively. 

\subsection{The spectral indices and the turnover frequency}

The GPS radio sources are a mixed group of galaxies and quasars,
with some remarkable differences between the two classes.
The histogram in Fig. 16a shows that the redshift distribution
is very different for galaxies and quasars. The galaxies have a typical
redshift of $\sim$ 0.5, and none has a redshift higher than 1.  For the galaxies 
without redshift information, we note that only 0316+161 has an optical 
magnitude slightly fainter than the galaxy 2128+048 at redshift 0.99 
(Table. 1).  Since these galaxies follow the Hubble diagram (O'Dea \etal\ 1996;
Snellen \etal\ 1996) it is unlikely they will be found at a redshift 
much higher than the others in the sample.
The quasars  are instead found at any redshift (we included the
galaxy 1404+286 (OQ208), which has a Seyfert 1 nucleus, in the quasar class)
with the majority at very high z (see also O'Dea 1990).

\begin{figure}
     \vbox{\psfig{figure=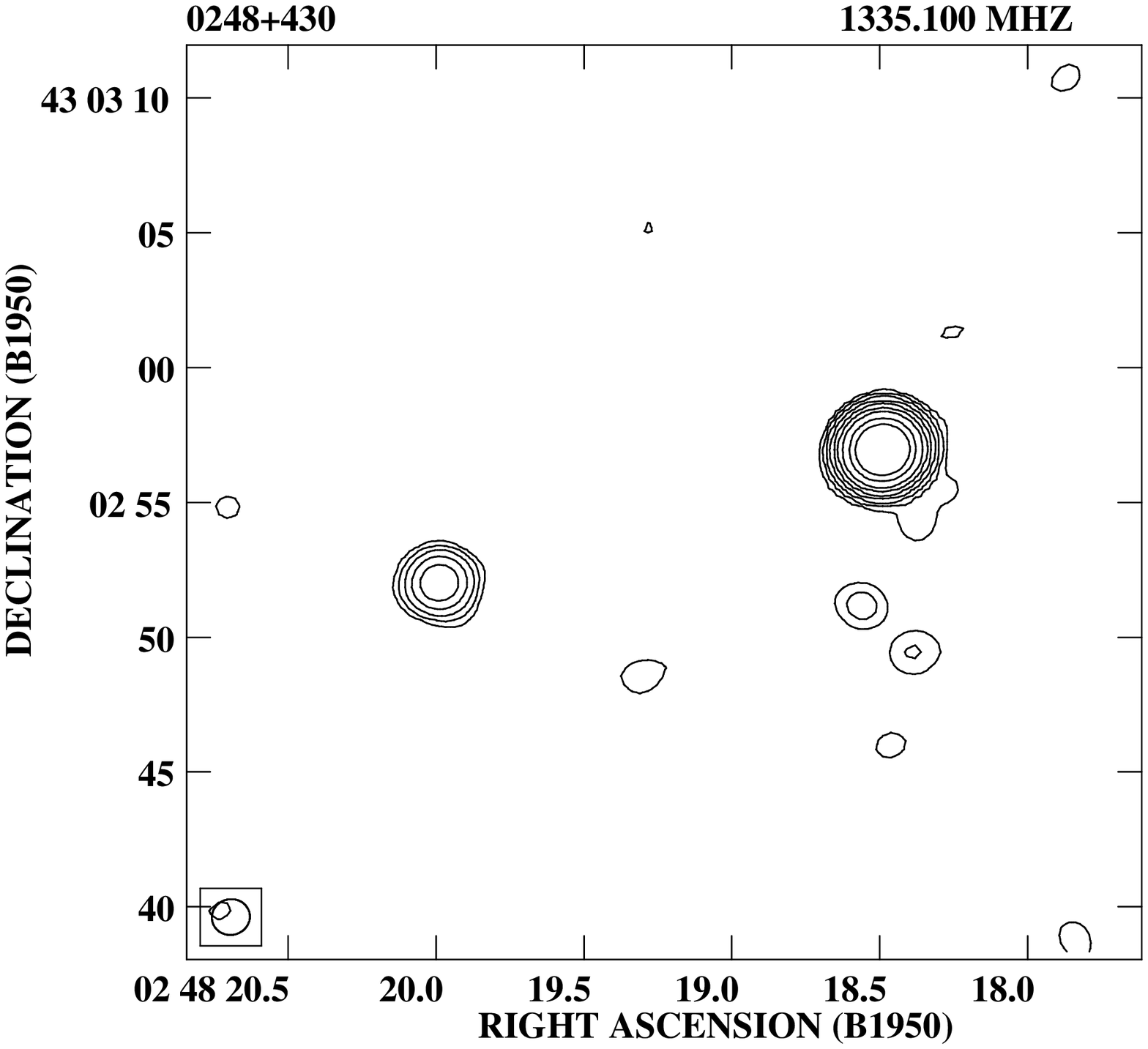,width=8cm,height=9cm}}\par
\caption[]{$0248+430$ at 1.35 GHz. The restoring beam is
1.41 $\times$ 1.32 arcsec in P.A. $-78^\circ$. The r.m.s. noise on the image is
0.2 mJy. The peak flux is 810 mJy/beam.
The contour levels for all the images are -3, 3, 6, 12, 25, 50, 100, 200,
500, 1000 $\times$ the r.m.s noise}
\end{figure}

\begin{figure}
     \vbox{\psfig{figure=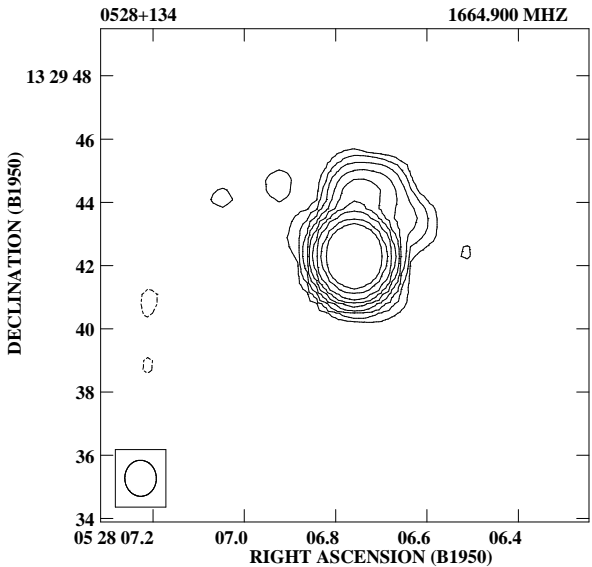,width=8cm,height=9cm}}\par
\caption[]{$0528+134$ at 1.66 GHz. The restoring beam is
1.14 $\times$ 1.01 arcsec in P.A. $+2^\circ$. The r.m.s. noise on the image is
0.25 mJy. The peak flux is 2088 mJy/beam.
}
\end{figure}

\begin{figure}
     \vbox{\psfig{figure=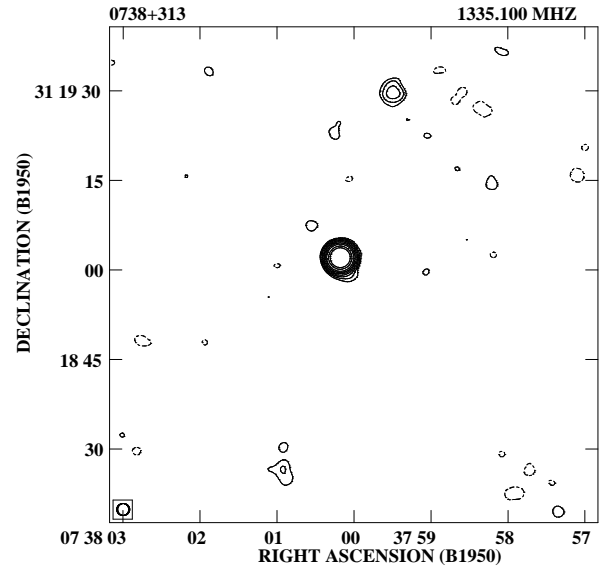,width=8cm,height=9cm}}\par
\caption[]{$0738+313$ at 1.33 GHz. The restoring beam is
2 $\times$ 2 arcsec. The r.m.s. noise on the image is
0.3 mJy. The peak flux is 1943 mJy.
}
\end{figure}

\begin{figure}
     \vbox{\psfig{figure=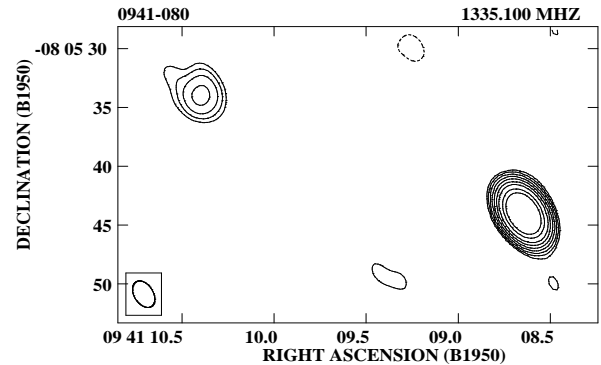,width=8cm,height=6cm}}\par
\vskip -1cm
\caption[]{$0941-080$ at 1.33 GHz. The restoring beam is
2.44 $\times$ 1.54 arcsec in P.A. $+29^\circ$. The r.m.s. noise on the image is
0.5 mJy. The peak flux is 2088 mJy/beam.
}
\end{figure}

\begin{figure}
     \vbox{\psfig{figure=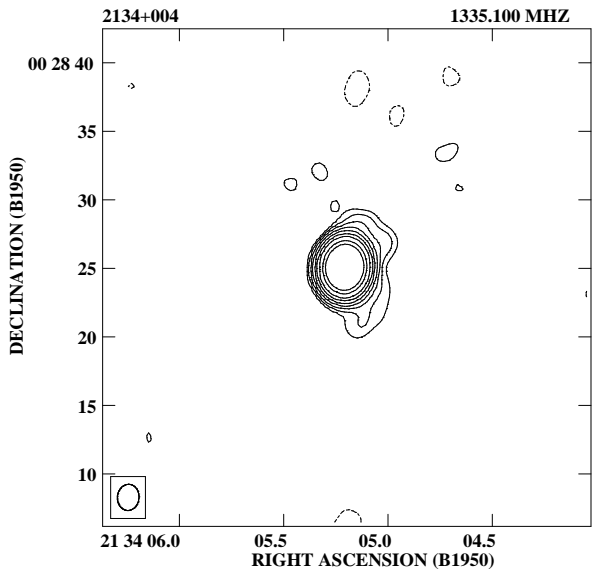,width=8cm,height=9cm}}\par
\caption[]{$2134+004$ at 1.33 GHz. The restoring beam is
1.91 $\times$ 1.55 arcsec in P.A. $-5^\circ$. The r.m.s. noise on the image is
0.4 mJy. The peak flux is 3249 mJy/beam.
}
\end{figure}

\begin{figure}
     \vbox{\psfig{figure=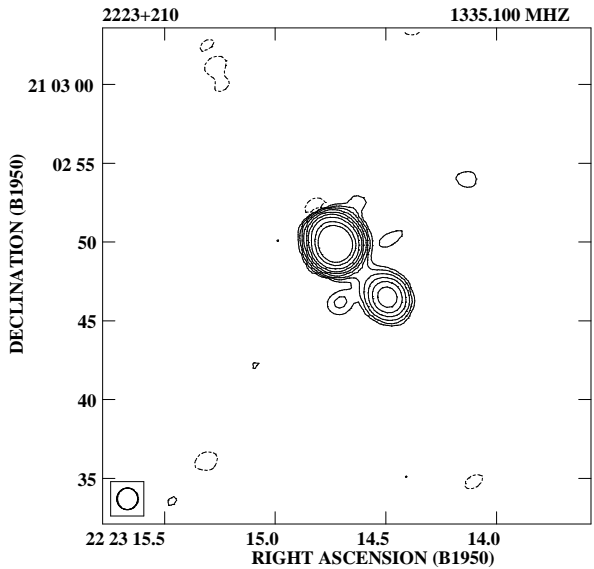,width=8cm,height=9cm}}\par
\caption[]{$2223+210$ at 1.33 GHz. The restoring beam is
1.37 $\times$ 1.32 arcsec in P.A. $-7^\circ$. The r.m.s. noise on the image is
0.3 mJy. The peak flux is 1663 mJy/beam.
}
\end{figure}

\begin{figure}
     \vbox{\psfig{figure=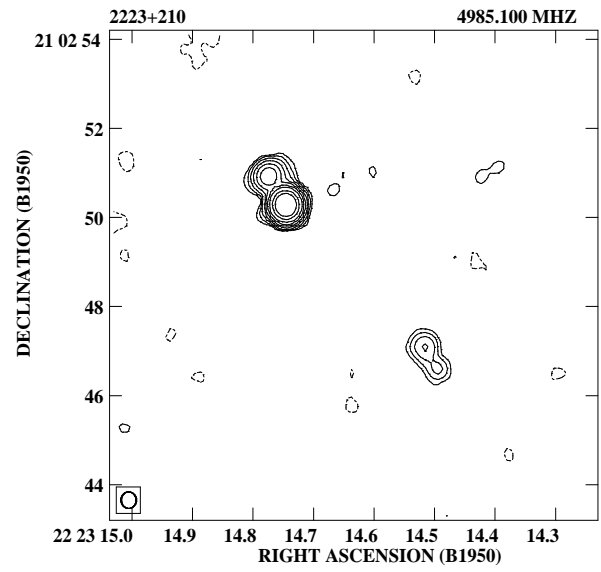,width=8cm,height=9cm}}\par
\caption[]{$2223+210$ at 5 GHz. The restoring beam is
0.37 $\times$ 0.35 arcsec in P.A. $+5^\circ$. The r.m.s. noise on the image is
0.3 mJy. The peak flux is 1034 mJy/beam.
}
\end{figure}

\begin{figure}
     \vbox{\psfig{figure=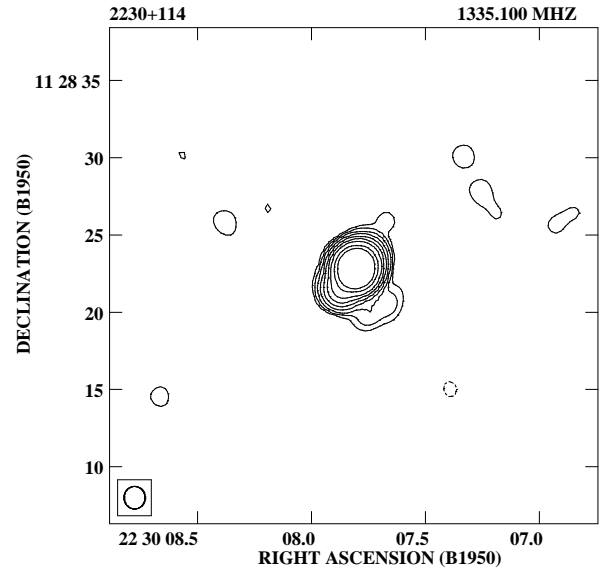,width=8cm,height=9cm}}\par
\caption[]{$2230+114$ at 1.33 GHz. The restoring beam is
1.46 $\times$ 1.36 arcsec in P.A. $+1^\circ$. The r.m.s. noise on the image is
0.8 mJy. The peak flux is 6444 mJy/beam.
}
\end{figure}

\begin{figure}
     \vbox{\psfig{figure=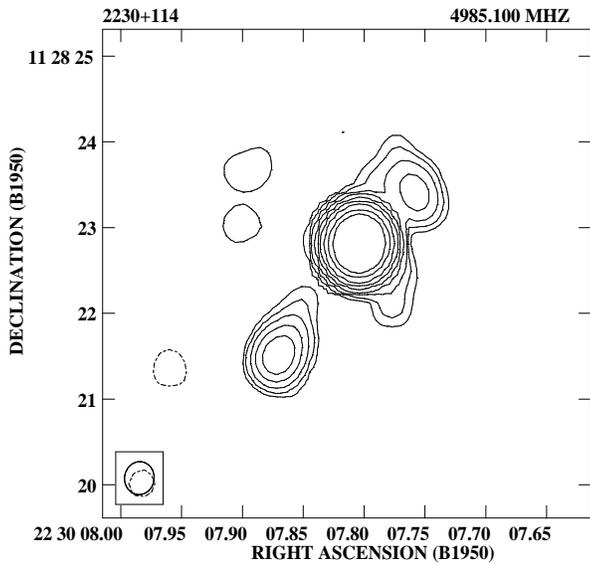,width=8cm,height=9cm}}\par
\caption[]{$2230+114$ at 5 GHz. The restoring beam is
0.38 $\times$ 0.36 arcsec in P.A. $-11^\circ$. The r.m.s. noise on the image is
0.5 mJy. The peak flux is 4137 mJy/beam.
}
\end{figure}

\begin{figure*}
     \vbox{\psfig{figure=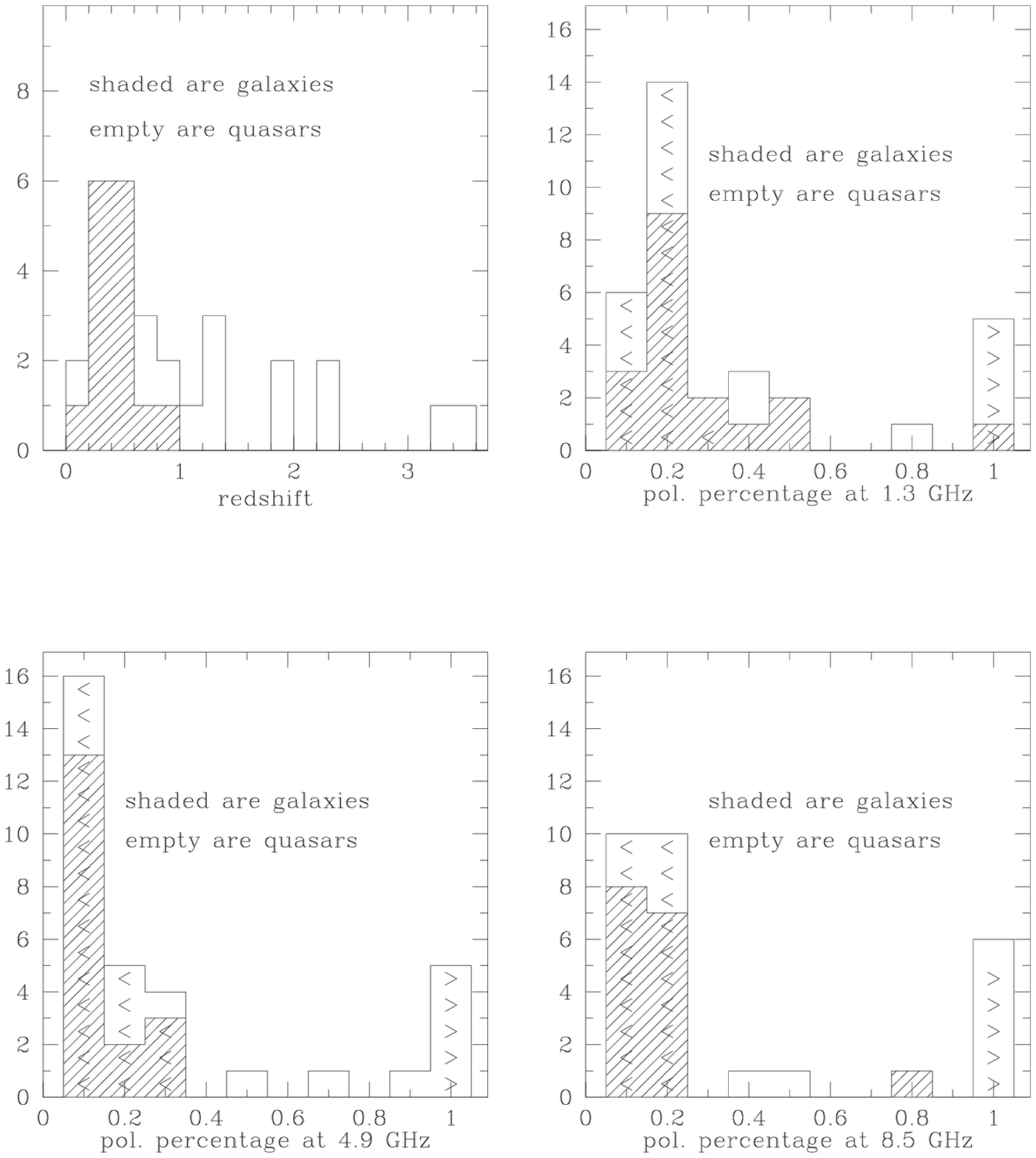,width=16.5cm,height=22cm}}\par
\vskip -6cm
\caption[]{Histograms for the complete sample:
a) redshift distribution; b) fractional polarization at 1.3 GHz; 
c) fractional polarization at 4.9 GHz; d) fractional polarization at 8.5 GHz }
\end{figure*}

The high frequency spectral index ranges from 0.5 (the limit set in 
the selection criteria) to 1.3 with the galaxies having perhaps slightly 
steeper values (but the 2 objects with the steepest spectral indices are quasars).
The low frequency spectral index ranges from -0.2 to -2.1
without any clear difference between galaxies and quasars,
but the higher values are biased since in several
objects the low frequency part of the spectrum is under-sampled
and the spectral index is calculated close to the turnover
frequency where it is likely to be flatter.  
Similar results were found by De Vries \etal\ (1997) though their poorer
frequency coverage resulted in a somewhat smaller range in
spectral index.

The quasars tend to peak at higher frequencies than the galaxies 
in both the rest frame and observed frame and  some quasars have a turnover 
frequency in their rest frame exceeding 10 GHz. 
This suggests that, on the assumption that the turnover is caused by synchrotron 
self-absorption, the GPS quasars are more compact than the galaxies. 
This effect has been also found by De Vries \etal\ (1997) in a bright
heterogeneous sample and by Snellen (1997) in a fainter sample.
In addition, VLBI images of several  sources belonging to the complete sample
show that quasars are more compact than galaxies, and in general exhibit 
 different morphologies (Stanghellini \etal\ 1997).

These results suggest that either GPS galaxies and GPS quasars are different
types of objects, or that beaming of compact components
plays a role in the quasars (see also O'Dea 1998).

\begin{table*}
\caption{Polarization for the complete sample
}
\begin{flushleft}
\begin{tabular}{rrrrrrrrrrrrrrr}
\hline
\noalign{\smallskip}
\hline
\noalign{\smallskip}
        &   1380  &&      1640   &&     4835   &&     4885&& && &&&\\
 name   &  $\%$  &   PA  &  $\%$   &  PA&    $\%$  &   PA &   $\%$  &   PA &   &   &   &&& \\
\noalign{\smallskip}
\hline\noalign{\smallskip}
0019$-$000&   0.5&  +0 &  0.3   &-2 &  $<$0.1&    &     $<$0.1& & & & &&&\\
0237$-$233&   2.0&  +7 &  1.2   &+4 &  4.0&  -35  & 4.0    &-35 & & & &&&\\
0500+019&  $<$0.2&     & $<$0.2   &   &  $<$0.1&    &     $<$0.1& & & & &&&\\
\noalign{\smallskip}
\hline
\noalign{\smallskip}
        &   1380&&        1630&&        4815&&        4865 &&       8435&&         8485&&&\\
 name   &  $\%$  &   PA  &  $\%$   &  PA&    $\%$  &   PA &   $\%$  &   PA &   $\%$ &    PA &    $\%$  &  PA&&\\
\noalign{\smallskip}
\hline
\noalign{\smallskip}
0108+388& $<$0.2&      &  0.5  &     & $<$0.2&&   $<$0.1&&       $<$0.1&       &  $<$0.1&     & &\\
0316+161&  0.5&  -63 &  0.7  & -60 & $<$0.1&&     $<$0.1&&       $<$0.1&       &  $<$0.1&    &&\\
0428+205&  0.4&  -59 &  0.6  & -58 & $<$0.1&&     $<$0.1&&       $<$0.1&       &  $<$0.1&   &&\\ 
0710+439& $<$0.1&      & $<$0.3  &     & $<$0.1&& $<$0.1&&       $<$0.1&       &  $<$0.1& &&\\
1031+567& $<$0.2&      & $<$0.2  &     & $<$0.3&& $<$0.2&&       $<$0.2&       &  $<$0.2&  && \\
1323+321&  0.3&  -35 &  0.4  & -01 & $<$0.1&&     $<$0.1&&        0.9&   +14 &   0.8&  +15&&\\  
1404+286&  0.4&  -50 & $<$0.3  & -23 & $<$0.2&&   $<$0.1&&       $<$0.2&       &  $<$0.2&&&\\
1518+047& $<$0.2&      & $<$0.2  &     & $<$0.3&& $<$0.1&&       $<$0.2&       &  $<$0.2& & & \\ 
1607+268& $<$0.3&      & $<$0.2  &     & $<$0.1&& $<$0.1&&       $<$0.1&       &  $<$0.1& &&\\
2008$-$068&  1.7&  -19 &  2.5  & -21 & $<$0.3&&     $<$0.3&&       $<$0.3&       &  $<$0.3& &  \\
2352+495& $<$0.2&      & $<$0.3  & -34 & $<$0.1&& $<$0.1&&       $<$0.1&       &  $<$0.1& &&\\
\noalign{\smallskip}
\hline
\noalign{\smallskip}
          &  1335&&        1665&&        4535 &&       4985&&        8085 &&        8465&&RM(rad/m$^2$)&\\
 name    & $\%$  &   PA &   $\%$   &  PA  &  $\%$ &    PA &   $\%$  &   PA &   $\%$ &    PA&     $\%$ &   PA&obs&rest\\
\noalign{\smallskip}
\hline
\noalign{\smallskip}
0248+430&  1.7&  +52? &  1.5  & +23  & 2.0&   -18 &  2.0 &  -27 &  0.5&   -40 &   0.4&  -44&131&703\\
0457+024& $<$0.1&      & $<$0.1  &      &$<$0.3&       &  0.3 &  +47 &  1.0&   +14 &   1.2&  +12&258&2954\\
0738+313& $<$0.1&      & $<$0.3  &      & 1.7&   +41 &  1.6 &  +67 &  3.0&   -2  &   3.2&  +2&-813&-2160\\
0742+103& $<$0.2&      & $<$0.2  &      &$<$0.1&       & $<$0.2 &      & $<$0.1&       &  $<$0.1& & & \\
0743$-$006&  0.4&  -76 &  0.5  & -7   &$<$0.2&       & $<$0.2 &      &  0.8&   +1  &   1.0&  -18& &\\
0941$-$080& $<$0.2&      & $<$0.1  &      &$<$0.1&       & $<$0.1 &      & $<$0.2&       &  $<$0.2&    & &\\
1117+146& $<$0.2&      & $<$0.1  &      &$<$0.2&       & $<$0.3 &      & $<$0.2&       &  $<$0.2&    && \\
1127$-$145&  4.3&  -84 &  3.9  & +61  & 2.6&   -26 &  2.8 &  -24 &  3.3&   -17 &   3.3&  -18 &-49&-234\\
1143$-$245&  1.9&  +72 &  2.1  & +80  & 0.7&   -18 &  0.7 &  -27 &  1.3&   -43 &  1.4 &  -46&146&1270\\
1245$-$197& $<$0.2&      & $<$0.1  &      &$<$0.1&       & $<$0.1 &      & $<$0.1&       & $<$0.1 &  && \\
1345+125& $<$0.2&      & $<$0.1  &      &$<$0.1&       & $<$0.2 &      & $<$0.1&       & $<$0.1 & & & \\ 
1358+624& $<$0.2&      & $<$0.1  &      &$<$0.1&       & $<$0.3 &      & $<$0.2&       & $<$0.2 & &&\\
1442+101&  0.8&  -74 &  0.8  & -78  & 2.0&   +60 &  1.7 &  +64 &  1.4&   +82 &  1.6 &  +78&-116&-2395\\
1600+335& $<$0.2&      & $<$0.1  &      &$<$0.1&       & $<$0.1 &      & $<$0.1&       & $<$0.1 & &  & \\
2126$-$158& $<$0.2&      & $<$0.2  &      &$<$0.1&       & $<$0.2 &      & $<$0.1&    & $<$0.2 & & &\\ 
2128+048& $<$0.1&      & $<$0.1  &      &$<$0.1&       & $<$0.1 &      & $<$0.2&       & $<$0.2 & & &\\
2134+004& $<$0.1&      & $<$0.1  &      & 0.9&   +60 &  0.9 &  +43 &  0.5&   -4  &  0.5 &  +0&349&3008\\
2210+016& $<$0.1&      & $<$0.1  &      &$<$0.1&       & $<$0.1 &      & $<$0.2&       & $<$0.2 &   && \\
2342+821& $<$0.2&      & $<$0.1  &      &$<$0.1&       &  0.5 &  -63 & &&&&&\\ 
\noalign{\smallskip}
\hline
\noalign{\smallskip}
\end{tabular}
\end{flushleft}
\end{table*}

\begin{table*}
\caption{Polarization of additional objects
}
\begin{flushleft}
\begin{tabular}{rrrrrrrrrrrrrrr}
\hline
\noalign{\smallskip}
\hline
\noalign{\smallskip}
         &  1335&&        1665&&        4535&&        4985&&        8085&&         8465&&RM(rad/m$^2$)&\\
 name   &  $\%$  &   PA  &  $\%$  &   PA   & $\%$     &PA   & $\%$     &PA    &$\%$     &PA     &$\%$    &PA&obs&rest\\
\noalign{\smallskip}
\hline
\noalign{\smallskip}
0404+768& $<$0.2&       &$<$0.1 &       &$<$0.1&     &   $<$0.3&      &       &     &       &    &&\\
0440$-$003&  0.9&  +49  & 0.9 & -27   & 2.3&  +61&    2.1&   +58&    2.6&  +55&     2.6 & +56&28&\\
0528+134&  0.6&  -56  &$<$0.3 &       & 2.2&  -31&    2.6&   -33&    3.9& -39 &    4.0  & -41&52&\\
2149+056& $<$0.2&       &$<$0.2 &       &$<$0.1&     &   $<$0.1&      &  $<$0.2 &     &   $<$0.2  &   && \\     
2223+210& 13.3&  -28  &10.5 & -82   &10.0&  -64&    9.1&   -58&       &     &        &-123&-1077\\
2230+114&  2.5&  -87  & 2.6 & -33   & 1.2&  +61&    1.5&   +75&       &     &       &-59&-244\\
\noalign{\smallskip}
\hline
\noalign{\smallskip}
        &   1380& &       1630&  &      4815&&        4865 &&       8435&&         8485&&RM(rad/m$^2$)&\\
 name   &  $\%$  &   PA  &  $\%$  &   PA   & $\%$     &PA   & $\%$     &PA    &$\%$     &PA     &$\%$    &PA&obs&rest\\
\noalign{\smallskip}
\hline
\noalign{\smallskip}
0026+346&  0.4&  -45  & 0.6 & -43    &$<$0.2&     &   $<$0.2&     &   $<$0.1&     & $<$0.1&&&\\
0201+113&  0.7&  -48  & 0.9 & -38    & 1.4&  +33&    1.4&  +33&    0.5&  +4 &  0.5&  +5&131&2724\\
0552+398&  0.4&  -52  & 1.0 & +78    & 0.8&  +43&    0.6&  +53&    0.8&  +66&  0.8&  +66&-1344&-15218\\
0703+468& $<$0.1&       &$<$0.2 &        &$<$0.2&     &   $<$0.2&     &   $<$0.3&     &  $<$0.3&&&\\
0711+356&  1.3&  +5   & 1.0 & -33    & 1.8&  +83&    1.8&  +80&    1.9&  +76&  2.0& +79&40&275\\
0904+039& $<$0.2&       &$<$0.2 &        &$<$0.5&     &   $<$0.5&     &   $<$0.9&     &  $<$0.9&&&\\
0914+114& $<$0.2&       &$<$0.2 &        &$<$0.8&     &   $<$0.8&     &   $<$2.0&     &  $<$2.0&&&\\
1543+005& $<$0.1&       &$<$0.2 &        &$<$0.2&     &   $<$0.2&     &   $<$0.2&     &  $<$0.1&&&\\
1732+094& $<$0.2&       &$<$0.3 &        & 1.5&  +16&    1.4&  -1 &   $<$0.3&     &  $<$0.3&&&\\
2015+657&  4.5&  -53  & 3.6 & -71    & 3.2&  +54&    3.1&  +55&    4.3&  +51&  4.3&  +51&30&\\
2021+614&  0.6&  -53  & 1.0 & -51    &$<$0.2&     &   $<$0.2&     &   $<$0.1&     &  $<$0.1&&&\\
2050+364& $<$0.2&       & 0.4 & -77    &$<$0.2&     &   $<$0.1&     &   $<$0.1&     &  $<$0.1&&&\\
2137+209& $<$0.2&       &$<$0.1 &        &$<$0.2&     &   $<$0.2&     &    0.4&  +68&  $<$0.3&&&\\
2337+264& $<$0.2&       &$<$0.3 &        &$<$0.3&     &   $<$0.2&     &   $<$0.2&     &  $<$0.2&&&\\
\noalign{\smallskip}
\hline
\noalign{\smallskip}
\end{tabular}
\end{flushleft}
\end{table*}


\subsection{Extended emission}

We have detected extended emission (both diffuse and compact) close to the 
compact radio source in some cases at 21 cm. In the remaining sources, 
our upper limits on extended emission is typically 1 mJy/beam at 21 cm. 

0248+430 (Figure 7) has a compact emitting region 15 arcsec east of
the main component and a hint of weak emission 5 arcsec to the south.
0528+134 is resolved, showing an extension in the NW direction (Figure 8).
Murphy \etal\ (1993) present an image of 0738+313 at 20 cm showing
2 emitting regions resembling 2 weak hot-spots and lobes on the
opposite sides of the dominant component. In our image (Figure 9) 
these 2 weak components are almost completely resolved out and only
a hint of emission has been detected 30 arcsec north and south
of the compact region. 0941-080 shows a slightly resolved secondary
component 20 arcsec east of the main one (Figure 10). 2134+004 has
 very weak and diffuse emission around the strong compact component
(Figure 11).
2223+210 has a secondary component 4 arcsec away from the main one
in the SW direction in our image at 1.35 GHz (Figure 12); the main component
itself is resolved in a core-jet structure 
oriented  NE with a possible counter jet
in the image at 5 GHz (Figure 13). 2230+114 at 1.35 GHz (Figure  14) 
shows an elongated structure
in the NW-SE direction with a hint of emission bending to SW, while in the
4.9 GHz image (Figure 15) the elongated structure turns out to be a core-jet
structure  with  the possible presence of a counter jet.

Stanghellini \etal\ (1990) report several cases of extended emission
around GPS radio sources.
Of the objects presented here showing extended  emission, 
0528+134 and 2223+210 are not true GPS objects (see also section 4.4 
for a discussion of the case of 0528+134).
In a couple of sources (0248+430, 0941-080 both belonging to the complete 
sample) 
it is difficult to say whether the secondary emission
is related to the GPS radio source and further observations
are probably needed. The extended emission found around
0738+313, 2134+004, and 2230+114 (the first 2 objects belong to the
complete sample) is likely to be related to
the GPS object. In the complete sample, 0108+388 is known to have
extended emission, so there are 3 to 5 objects out of 33  with known
extended emission so far. This percentage of 9 to 15$\%$ is slightly smaller 
but consistent with that
previously claimed by Stanghellini \etal\ (1990). 
It is clear that the vast majority ($\sim 90\%$)
of the GPS sources appear to be truly isolated and have no emission
beyond the kpc scale at the current limits.  

\subsection{Variability}

Waltman \etal\ (1991) presented monitoring observations at
2.7 and 8.1 GHz for several GPS sources covering the time range
from 1979 to 1988.
Some sources as 0237-233, 1245-197, 1345+125 were found to be 
very stable in flux density. Others were found to be variable: 
0552+398 shows a variation
of $\sim$ 30 $\%$ at 8.1 GHz. 2134+004 has a variability of about 15-20$\%$
at 8.1 GHz. 2352+495 has a variability below 10$\%$ at 8.1 GHz. Also
0742+103 is slightly variable. 

Wehrle \etal\ (1992) also report variability for some GPS objects
in the time range 1985-1991
at 4.8, 8, and 14.5 GHz,
from the University of Michigan Radio Astronomy Observatory monitoring
program for 
several sources, some of which are GPS objects. 0552+398 shows an increase
in flux density exceeding 50$\%$ at 8.4 and 14.5 GHz, 1127-145 shows
a quasi periodical variability of approximately 1 Jy at all the 3 frequencies.
The source 2230+114 also shows a rather remarkable flux density variability at all
the 3 frequencies with an amplitude of 0.5-1 Jy and is a well known low frequency 
variable source (Bondi \etal\ 1996).  The variability of 1404+286
has been discussed by Stanghellini \etal\ (1996).

In conclusion,  we find that some  GPS sources (mainly quasars) show 
mild to high flux density variability at cm and mm wavelengths. However,
without uniform monitoring of the complete sample it will not be possible 
to determine how common this variability is. 
We also note a couple sources where the spectral shape is variable and at
some times the spectrum was peaked, and at other times it was not,
0528+134, the well known gamma-ray source (Mukherjee \etal\ 1996), and 0201+110. 
Both of these sources show a rather flat spectrum in the
VLA observations from the second or the third session
and they would be very easily discarded as GPS radio sources.
But 0528+134 has been included in the class of GPS radio sources
because of its GPS-like spectrum from the literature (O'Dea \etal\ 1991), 
and 0201+113 really shows a convex spectrum in the  data
in the first VLA session published in O'Dea \etal\  (1990).
This behavior is not surprising in highly variable radio sources
as we may well expect that the presence of new radio components will change 
the spectral shape.  Thus, there are sources which show a peaked
spectrum only part of the time. This aspect of the GPS phenomenon  deserves more 
attention as it could be related to the remarkably different properties found 
between GPS galaxies and (some?) GPS quasars.

\subsection{Polarization}

Due to the low level of observed polarization, the errors are dominated
by the r.m.s noise on the polarization images (typically 0.5 mJy) and by 
contamination from residual unpolarized emission (estimated about 0.2-0.3$\%$ of the 
total flux density).
The error in the polarized flux may then be calculated
as $\sigma _P=\sqrt {0.5^2 + (0.003\times S_{mJy})^2}mJy$.
We considered any measurement of the fractional 
polarization below 0.3$\%$ to be an upper limit (Tables 6 and 7).
The errors in the position angle have a systematic
contribution due to the uncertainty in the determination
of the position angle of the calibrator (3C286), assumed to
be around 2-3 degrees. This is the dominant
contribution in the position angle error for most of the objects with
detected polarized flux density.

In Fig. 16 we show the histograms of the fractional 
of polarization (mostly upper limits) for the complete sample 
at 1.3, 4.9 and 8.5 GHz.
The fractional polarization is in general low at all the frequencies.
Only a few quasars have a fractional polarization above 1 $\%$ at
4.9 or 8.5 GHz.

When polarized emission has been detected we attempted a linear fit to
the polarization angles versus the squared observed wavelength,
as is expected from the Faraday effect on polarized 
radiation propagating through a magnetized and ionized medium.

The low level or even the lack of detection of polarized
flux limited us to only a few sources  (all quasars). The fits are generally rather 
good and are given in Table 6 and 7, and 
in Figures 17 and 18. 
Due to the better frequency coverage in the present observations, our estimated 
rotation measures supersede those reported by O'Dea \etal\ (1990), though we cannot 
rule out that some of the difference is due to variability. 
We find Faraday rotation measures in the rest frame above 1000 rad/m$^2$ for
5 quasars of the complete sample. We also found a very high
value ($>10^4$ rad/m$^2$) for 0552+398 which does not belong to the complete sample
but has a GPS shape.

Sometimes the frequencies which give a good fit include those close to the 
turnover, and in the case of 0552+398 are all below the turnover.
This implies that the region emitting the polarized emission is different 
from that responsible for the optically thick emission or that the turnover is not 
caused by synchrotron self absorption.

\subsection{Summary and conclusions}

We have presented a bright flux-density-limited complete sample of 
33 GPS radio sources selected on the basis of  their peaked radio spectra.  
The sample selection was based on observations with the VLA, WSRT, and other 
instruments.  Additional GPS sources not belonging to the complete sample have 
also been observed. We present our results on the polarization and radio spectrum. 
We found remarkable differences in the properties of  quasars and 
galaxies, the latter having lower turnover frequencies,  mostly undetectable
polarization and lower redshifts.

In the few objects where polarization has been detected at many 
frequencies, the Faraday rotation measure in the rest frame often
exceeds 1000 rad/m$^2$.

In about 10\% of the sources we detect weak diffuse extended emission.
In the remaining $\sim 90\%$ any extended emission has a peak surface
brightness is less than about 1 mJy/beam at 21 cm. 

In a following paper we will discuss the implications of the
properties of the complete sample in the framework of the
scenarios proposed to explain the existence of the GPS radio sources.  

\acknowledgements{
We thank Ger de Bruyn for advice on the reduction of the WSRT observations
and Wim De Vries for comments on the manuscript.
C.S. wishes to thank the STScI Collaborative Visitor Program 
for providing support for his visits.
The VLA is operated by the U.S. National Radio Astronomy Observatory
which is operated by Associated Universities, Inc., under cooperative
agreement with the National Science Foundation.
The Westerbork Synthesis Radio Telescope is operated by the Netherlands
Foundation for Research in Astronomy (NFRA) which is financially supported
by the Netherlands organization for scientific research (NWO) in the Hague.
We have made use of the NASA/IPAC Extragalactic Database, operated by the
Jet Propulsion Laboratory, California Institute of Technology,
under contract with NASA.}

\end{document}